\documentclass[12pt]{article}
\newif\ifams\amsfalse                                                    
\amstrue                                                                 
\ifams                                                                   
 \usepackage{amsfonts}                                                   
\fi                                                                      
\newif\iffigs\figsfalse                                                  
\newif\ifdraft\draftfalse
\newif\ifinter\interfalse
\intertrue
\ifdraft\else\interfalse\fi
\ifinter
 \else
\fi
\ifinter
  \def\secl#1{\nopagebreak\marginpar{\vspace{-6mm}\scriptsize #1}\label{#1}}
  \def\beql#1{\marginpar{\vspace{4mm}\scriptsize #1}
              \nopagebreak\begin{equation}\label{#1}}
  \def\ftl#1#2{\footnote{\label{#1}[#1] #2}}
  
  \def\bibl#1{\marginpar{\vspace{4mm}\scriptsize #1}\nopagebreak\bibitem{#1}}
 \else
  \def\secl#1{\label{#1}}
  \def\beql#1{\begin{equation}\label{#1}}
  \def\ftl#1#2{\footnote{\label{#1}#2}}
  
  \def\bibl#1{\bibitem{#1}}
\fi

\def\tempnote#1%
  {\hspace{1mm}\newline\noindent\begin{tabular}[t]{|p{15.5cm}|}
     \hline \rule{0mm}{2.5ex}#1 \\ \hline
    \end{tabular}\newline\noindent}

\def\draftnote#1%
  {\ifdraft
    \hspace{1mm}\newline\noindent\begin{tabular}[t]{|p{15.5cm}|}
     \hline \rule{0mm}{2.5ex} \underline{DRAFT NOTE}: #1 \\ \hline
    \end{tabular}\newline\noindent
   \fi}

\def\internote#1%
  {\ifinter
    \hspace{1mm}\newline\noindent\begin{tabular}[t]{|p{15.5cm}|}
     \hline \rule{0mm}{2.5ex} \underline{Internal Note}: #1 \\ \hline
    \end{tabular}\newline\noindent
   \fi}

\def\multdn{$\downarrow\downarrow\downarrow\downarrow\downarrow$}
\def\beginsup%
 {\noindent\begin{tabular}[t]{|c|}
   \hline  Supplements\\
   \multdn\multdn\multdn\multdn\multdn\multdn
   \multdn\multdn\multdn\multdn\multdn\multdn \\ \hline
  \end{tabular}}

\def\multup{$\uparrow\uparrow\uparrow\uparrow\uparrow$}
\def\endsup%
 {\noindent\begin{tabular}[t]{|c|}
   \hline \multup\multup\multup\multup\multup\multup
          \multup\multup\multup\multup\multup\multup \\
   Supplements\\ \hline
  \end{tabular}}

\iffigs
 \input epsf
 
\else
\fi
\newif\ifappend\appendfalse
\appendtrue
\ifappend
 \newcommand{\newsection}[1]{
  \vspace{7mm} \pagebreak[3]
  \refstepcounter{section}
  \setcounter{equation}{0}
  \message{(\thesection. #1)}
  \addcontentsline{toc}{section}{
   \protect\numberline{\thesection}{\hs\hs\boldmath #1}}
  \begin{flushleft}
   {\large\bf\boldmath \thesection. #1}
  \end{flushleft}
  \nopagebreak}

\else
 \newcommand{\newsection}[1]{\section{#1}}
\fi

\newcommand{\newpar}[1]{
 \vspace{3mm}
 \noindent{\bf\boldmath #1}
 \nopagebreak\vspace{2mm}\nopagebreak}

\def\al{\alpha}
\def\bt{\beta}
                \def\Gm{\Gamma}
\def\dl{\delta}                \def\Dl{\Delta}

\def\lm{\lambda}               \def\Lm{\Lambda}

               \def\Om{\Omega}
\def\sg{\sigma}               \def\Sg{\Sigma}

\def\Ac{\mbox{\protect$\cal A$}}
\def\Bc{\mbox{\protect$\cal B$}}
\def\Cc{\mbox{\protect$\cal C$}}

\def\Fc{\mbox{\protect$\cal F$}}
\def\Mc{\mbox{\protect$\cal M$}}
\def\Nc{\mbox{\protect$\cal N$}}

\def\Rc{\mbox{\protect$\cal R$}}
\def\Uc{\mbox{\protect$\cal U$}}

\ifams
 \def\bbl#1{{\mathbb #1}}
\else
 \def\bbl#1{{\bf #1}}
\fi
\def\CC{\bbl{C}}

\def\RR{\bbl{R}}
\def\ZZ{\bbl{Z}}

\def\tr{{\rm tr}}

\def\beq{\begin{equation}}
\def\eeq{\end{equation}}
\def\hs{\hspace{2mm}}
\def\hsc{\hspace{2mm},\hspace{5mm}}
\def\nl{\protect\newline}
\def\nlb{\protect\newline $\bullet$ }
\def\nls{\protect\newline $\star$ }
\def\ie{{\em i.e.}}
\def\eg{{\em e.g.}}
\def\etal{{\em et.\ al.\ }}

\def\pt{\partial}
\def\goto{\rightarrow}
\def\Goto{\hspace{5mm}\Rightarrow\hspace{5mm}}

\def\dg{^{\dagger}}

\def\vev#1{\left\langle#1\right\rangle}

\def\rec#1{{\raise 0.4ex \hbox{$\scriptstyle {\frac{1}{#1}}$}}}

\def\half{{\raise 0.4ex \hbox{$\scriptstyle {1 \over 2}$}}}

\def\hepth#1{{\tt hep-th/#1}}
\def\ATMP#1#2#3{{\it Adv. Theor. Math. Phys.} {\bf #1} (#2) #3}

\def\IJTP#1#2#3{{\it Int. J. Theor. Phys.} {\bf #1} (#2) #3} 
\def\IJMP#1#2#3{{\it Int. J. Mod. Phys.} {\bf #1} (#2) #3} 
\def\JHEP#1#2#3{{\it JHEP} {\bf #1} (#2) #3}
\def\FP#1#2#3{{\it Fortsch. Phys.} {\bf #1} (#2) #3} 
\def\NPB#1#2#3{{\it Nucl. Phys.} {\bf B#1} (#2) #3}
\def\NPPS#1#2#3{{\it Nucl. Phys.} {\bf #1} {\it (Proc. Suppl.)} (#2) #3}
\def\PLA#1#2#3{{\it Phys. Lett.} {\bf A#1} (#2) #3}
\def\PLB#1#2#3{{\it Phys. Lett.} {\bf B#1} (#2) #3}
\def\PRL#1#2#3{{\it Phys. Rev. Lett.} {\bf #1} (#2) #3}
\def\PRV#1#2#3{{\it Phys. Rev.} {\bf #1} (#2) #3}

\def\RMP#1#2#3{{\it Rev. Mod. Phys.} {\bf #1} (#2) #3} 

\def\WCP{\bbl{W}\bbl{C}\bbl{P}}
\begin{document}


\begin{titlepage}

\begin{flushright}
EFI-2000-1\\ {\tt hep-th/0001054}\\
\end{flushright}

\ifinter \vspace{-10mm} \else \vspace{5mm} \fi

\begin{center}
\LARGE {\bf\boldmath Holography, Singularities on Orbifolds \\
 and 4D $N=2$ SQCD} \\

\ifinter \vspace{5mm} \else \vspace{10mm} \fi

\large Oskar Pelc \normalsize 
\vspace{5mm}

{\em Enrico Fermi Institute\\ University of Chicago \\
5640 S. Ellis Ave. Chicago, IL 60037, USA} \\
E-mail: {\tt oskar@theory.uchicago.edu}

\vspace{5mm}
\ifinter \else \vspace{5mm} \fi
\end{center}

\ifinter \else \vspace{10mm} \fi

\begin{center}\bf Abstract\end{center}
\begin{quote}
Type II string theory compactified on a Calabi-Yau manifold, with a
singularity modeled by a hypersurface in an orbifold, is considered.
In the limit of vanishing string coupling, one expects a non gravitational
theory concentrated at the singularity. It is proposed that this theory is
holographicly dual to a family of ``non-critical'' superstring vacua,
generalizing a previous proposal for hypersurfaces in flat space.
It is argued that a class of such singularities is relevant for the
study of non-trivial IR fixed points that appear in the moduli space of
four-dimensional $N=2$ SQCD: $SU(N_c)$ gauge theory with
matter in the fundamental representation. This includes the origin in the
moduli space of the $SU(N_c)$ gauge theory with $N_f=2N_c$ fundamentals.
The 4D IR fixed points are studied using the anti-holographic description
and the results agree with information available from gauge theory.
\end{quote}


%
\internote{BEFORE SUBMISSION:
\nl Print: Preprint Nr.; Acknowledgments; Spell Check; Update References
\nl Source: draft flag; $\{\}$From; line overflow}

\end{titlepage}

\tableofcontents



\newsection{Introduction}

Duality between theories that relates strong and weak coupling is an
important tool in the study of strongly coupled systems. The holographic
duality, discovered in the  study of branes in string/M theory
\cite{Malda9711} (for a review, see \cite{AGMOO9905}), is such a duality.
It relates the dynamics of a brane system, in a situation in which
it decouples from the bulk dynamics, to string/M theory with the background
being the near-horizon geometry of that brane system.
\internote{
\nlb For the branes of M theory, this background includes an AdS factor.
In type II string theory, the same is true (in an appropriate conformal
frame) for the D branes.
\nlb For NS5 branes, the near horizon geometry includes
a dilaton that depends linearly on the radial direction (related to the
distance from the brane.}

A similar holographic relation was suggested in \cite{GKP9907}
(and  studied further in \cite{GK9909}\cite{GK9911}).
The system considered is type II string theory on
\beql{CYb} R^{d-1,1}\times X^{2n} \hs, \eeq
where $X^{2n}$ is a singular Calabi-Yau manifold. In the limit of vanishing
string coupling, the only non-trivial dynamics is of modes localized near the
singularity, defining a $d$-dimensional theory without gravity.
In \cite{GKP9907}, an ``anti-holographic'' formulation of this theory was
proposed: type II string theory in a background of the form
\beql{LDb} \RR^{d-1,1}\times\RR_\phi\times\Nc \hs, \eeq
where $\RR_\phi$ corresponds to the radial direction (distance from the
singularity) and the dilaton varies linearly along this direction.
These vacua were first studied in \cite{KS90} (for a review, see
\cite{Kutasov9110});
that they should exhibit holography, was argued in \cite{ABKS9808}%
\footnote{The relevance of a background with an $\RR_\phi$ factor as above
to the dynamics at singularities in CY compactification was also suggested,
from different points of view, in \cite{GV9506} and \cite{Witten9507}.}.
In the backgrounds studied in \cite{GKP9907}, the manifold $X^{2n}$ in
(\ref{CYb}) is a hypersurface $W(z_1,\ldots,z_{n+1})=0$ in flat space
$\CC^{n+1}$ and it has an isolated singular point.
In the present work, this study of the holographic duality is extended to
hypersurfaces in a supersymmetry-preserving orbifold $\CC^{n+1}/\Gm$
(where $\Gm$ is a finite group of unitary transformations). 

For a large class of manifolds $X^{2n}$, the background (\ref{CYb}) is T-dual
to a configuration with NS5 branes, so the $d$-dimensional theory describes
the decoupled dynamics of these branes.
For $d=6$, the T-duality leads to flat NS5 branes
\cite{OV9511}\cite{Kutasov9512}\cite{GHM9708} (see also \cite{KKL9308})
and the background (\ref{CYb}) is the near-horizon geometry of these branes
\cite{CHS9112}, $\RR_\phi$ being the radial direction.
For $d=4,2$, one obtains a single NS5 brane wrapped on a manifold $\Sg$
\cite{KLMVW9604}
and the singular point in $X^{2n}$ transforms to a singular point in $\Sg$.
In these cases, the background (\ref{LDb}) is identified with
the near-horizon geometry of this NS5 brane, in the singular region.

The dynamics of a type IIA NS5 brane wrapped on a Riemann surface $\Sg$
is relevant for the study of the low energy dynamics of 4D $N=2$
supersymmetric gauge theories.
This was shown, by a chain of dualities, starting with a
background where the gauge theory is seen perturbatively -- either in
heterotic string theory (as reviewed in \cite{Engineer}) or in brane
configurations in type IIA string theory (as reviewed in \cite{GK9802}).
The surface $\Sg$ is then identified with the {\em Seiberg-Witten} (SW) curve 
\cite{SW9407}\cite{SW9408},
which encodes information about the IR dynamics of the gauge theory.

Of particular interest are points in the moduli space, in which the dynamics
remains non-trivial in the IR limit.
They define interacting 4D $N=2$ superconformal field theories (SCFT's).
These points correspond to singular SW curves.
This suggests that such a SCFT is realized in string theory by a type
IIA NS5 branes wrapped on a singular surface $\Sg$ and, by T-duality,
in type IIB string theory on $R^{3,1}\times X^6$, with a singular $X^6$.
The proposed holographic duality then provides an additional framework in
which this SCFT can be studied.
One family of such SCFT's was identified \cite{AD9505} in SYM $SU(N_c)$
gauge theory, at points in the moduli space with mutually-non-local massless
dyons.
These points are indeed realized by singular hypersurfaces in flat space
and were considered, using the holographic duality, in \cite{GKP9907}.
Here this relation is extended to
SQCD -- $SU(N_c)$ gauge theory with matter (``quarks'') in the fundamental
representation. 
It is argued that when there are $k$ massless quarks at the interacting
IR SCFT, this theory is realized by a background with a singular
hypersurface in a $\ZZ_k$ orbifold.
The orbifold represents the near-horizon region of $k$ KK monopoles,
and their frozen 6D gauge symmetry provides the global (``flavor'') symmetry
of the 4D gauge theory.
The non-trivial IR SCFT's in the moduli space of SQCD were studied, using
SW theory, in \cite{APSW9511},\cite{EHIY9603}.
Here these theories are studied using the proposed holographic duality.
Specifically, by considering deformations from such a theory, one identifies
vector superfields in it and obtains information about their dynamics.
All the results agree with what is known from field-theoretic considerations.
One family of these SCFT's corresponds to the origin in
the moduli space of $SU(N_c)$ gauge theories with $N_f=2N_c$ massless quarks.
These theories are continuously connected to a free theory (by varying the
gauge coupling). The implications of this relation are compared to the
results obtained from the holographic duality and, again, agreement is found.

The structure of this work is as follows: in the next section, the decoupling
limit is reviewed (mostly following \cite{GKP9907},\cite{GK9909}),
both for singularities and for NS5 branes.
The decoupled theory is defined and some general aspects of it are discussed.
These aspects are then demonstrated in the case of hypersurface singularities.
\internote{Elements not appearing in \cite{GKP9907},\cite{GK9909}:
\nlb The relation to the $El_s\goto0$ limit.
\nlb Conditions for the relation to NS5 branes.
\nlb Trivial decoupled dynamics of a single NS5 brane [with David].}%
In section \ref{s-orb}, a hypersurface singularity on an orbifold is
considered. The holographicly-dual pair is identified and properties
of the $d$-dimensional theory, as obtained from the dual formulations,
are discussed.
This information is used in section \ref{s-SGT}, where it is applied to
specific cases.
In section \ref{s-SGT}, we concentrate on configurations related to
4D $N=2$ SQCD. The realization of this theory in string
theory is reviewed, the anti-holographic description is identified and
used to study the gauge theory.
The results of this work are summarized in section \ref{s-sum}.

\newsection{The Decoupling Limit}
\secl{s-dec}
Consider type II string theory on
\[ R^{d-1,1}\times\bar{X}^{2n} \hs,  \]
where $\bar{X}^{2n}$ is a compact, $n$ (complex) dimensional Calabi-Yau
(CY) manifold, in the limit of vanishing string coupling $g_s\goto0$.
When $\bar{X}^{2n}$ is smooth, the dynamics is expected to becomes free
in this limit. If, on the other hand, $\bar{X}^{2n}$ has a singular
point $s$, one expects in the above limit a non-trivial dynamics%
\footnote{Evidence for such a non-trivial dynamics is provided by the
relation between geometrical singularities and NS5 branes, as described
below.}
of modes localized near $s$.
If $s$ is ``isolated'', \ie,  has a smooth neighborhood,
these modes will decouple from the other modes
of the string theory, which are either localized at other singular points
or propagate in the bulk of $\bar{X}^{2n}$. In particular, they will decouple
from gravity (which corresponds to bulk modes).
Therefore, the decoupled dynamics of the modes localized on the singular
point $s$ is described by a $d$-dimensional theory without gravity.
This theory is expected to have an anti-holographic description, as explained
in the introduction, and it is the subject of the present investigation.

To obtain a formulation of this theory, it will be useful to have a simpler
definition of it. This is indeed possible:
one can replace $\bar{X}^{2n}$ by a simpler, non-compact manifold $X^{2n}$,
that shares with $\bar{X}^{2n}$ the same singular region $S$
(\ie, the singular point $s$ and its smooth neighborhood)
and the decoupling implies that this change will have no effect on the
decoupled dynamics in $S$.
In fact, to define such a $d$-dimensional theory, one can start directly with
a non-compact manifold $X^{2n}$. This also allows one to consider
singularities that cannot be embedded in a compact CY manifold.
\internote{
\nl Isn't there a problem with the definition of string theory on
non-compact CY's?}

A generalization of the above procedure was suggested in \cite{GK9909}:
instead of considering the limit $g_s\goto0$ with fixed $X^{2n}$, one can
study a ``double-scaling'' limit, in which the CY manifold is also
modified. For this, one considers a deformation $X^{2n}_\mu$ of the above
singular manifold $X^{2n}$, parametrized by a deformation parameter $\mu$.
The double-scaling limit is
\beql{dbl-lim}
g_s\goto0 \hsc \mu\goto0 \hsc m=\frac{\mu^r}{g_s}=\hbox{const.} \hs,
\eeq
where $m$ is an appropriate ratio that corresponds to some physical quantity
in the decoupled theory. The dependence of the limit on this ratio means, in
particular, that the separate limits $\mu\goto0$ and $g_s\goto0$ do not
commute. The original procedure, with fixed $X^{2n}$, \ie, first taking
$\mu\goto0$ and then $g_s\goto0$, corresponds to $m=0$, while the opposite
order of limits corresponds to $m=\infty$.

There is another limit in which one obtains a theory without gravity
which is localized on a singularity:
the low energy limit $El_s\goto0$, where $l_s$ is the string length scale and
$E$ is the characteristic energy considered.
This limit was used, for example, to study the low energy dynamics of
4D $N=2$ supersymmetric gauge theories (see \cite{Engineer} for reviews).
The above two limits are, in fact, two aspects of the same decoupling
phenomenon:
at fixed finite $g_s$, one obtains an approximate decoupling of bulk modes
below some energy $E_d$, where $E_d l_s$ depends on $g_s$
(and the moduli of the CY manifold).
This becomes an exact decoupling in the extreme IR.
As $g_s$ is decreased (with fixed $l_s$), $E_d$ increases
(\eg, the gravitational interaction becomes weaker, since
the Planck length $l_p\sim g_s^{1/4}l_s$ decreases)
and at vanishing $g_s$, one obtains a decoupled theory at all energy scales.
\internote{Is there an estimate for $E_d$?}

\newpar{The Relation to NS5 branes}
\nl An important special case is when $X^{2n}$ has a $U(1)$ isometry
(acting holomorphicly), and the space $\Sg$ of fixed points of this
isometry is a (complex) sub-manifold of co-dimension $2$.
Performing T-duality along the $U(1)$ isometry, one obtains \cite{KLMVW9604}
a configuration with an NS5 brane wrapped on $\Sg$.
\internote{Proof:
\nls the transverse neighborhood of a fixed point $\sim\CC^2$
\nls analytic isometries with a fixed point $=$ $U(2)$
\nls nlg, $U(1)$ is diagonal
\nls no other fixed points for $1\neq g\in U(1)$ $\goto$ a unique solution.}
Therefore, in this case, the $d$-dimensional theory defined above describes
the decoupled dynamics of this NS5 brane. 
The simplest such case is a configuration of parallel KK monopoles.
The transverse space (multi Taub-NUT space; $n=2$) has a $U(1)$ isometry,
with fixed points at the location of the monopoles.
T-duality along this isometry replaces each KK monopole by an NS5 brane
at the same location
\cite{OV9511}\cite{Kutasov9512}\cite{GHM9708} (see also \cite{KKL9308}).
When the NS5 branes coincide, one indeed expects, in the $g_s\goto0$ limit,
a non-trivial dynamics on these branes \cite{Seiberg9705} (see also
\cite{DDV9604}\cite{DDV9704}\cite{BRS9704}).
This is because they are dual to decoupled D5 branes, where one obtains,
at low energy, a gauge theory with a non-vanishing gauge coupling.

When the KK monopoles are each at a different location, the geometry is
smooth everywhere and, therefore, in the decoupling limit, $g_s\goto0$, the
dynamics is free. The above T-duality now implies that the decoupled
dynamics of a single NS5 brane is free. It is natural to expect that
the same remains true when the NS5 brane is wrapped on a smooth
manifold. When $\Sg$ has a singularity, the same reasoning suggests that 
the non-trivial dynamics will be confined to the singular region%
\footnote{The author is grateful to D. Kutasov for a discussion about this
issue.}

\newpar{The Case of a Hypersurface Singularity}

We close this section, by demonstrating some of the issues described above
in the specific cases considered in \cite{GKP9907}\cite{GK9909}\cite{GK9911}
and the present work: singularities embedded in a non-compact hypersurface
\[ W(z_1,\ldots,z_{n+1})=0 \]
in $\CC^{n+1}$, or an orbifold of $\CC^{n+1}$.
The modes localized on the singularity include modes related to D-branes
wrapped (fully or partially) on vanishing cycles.
These define particles or extended objects in the $d$-dimensional theory.
In the double scaling limit discussed in
\cite{GK9909}\cite{GK9911}, the quantity $m$ held fixed (see eq.
(\ref{dbl-lim})) is (proportional to) the mass/tension of these objects
(this is explained in subsection \ref{s-pair}).

When $W$ has the form
\[ W=H(z_1,\ldots,z_{n-1})+uv \hs, \]
$X^{2n}$ has a $U(1)$ isometry
\[ u\goto\al u \hsc v\goto v/\al \hsc |\al|=1 \]
and the space $\Sg$ of fixed points is%
\footnote{For $n=2$, this describes the configuration of parallel KK
monopoles discussed above.}
$u=v=H=0$.
If the hypersurface is in $\CC^{n+1}$, T-duality along the $U(1)$ isometry
gives \cite{KLMVW9604} an NS5 brane wrapped on $\Sg$ ($H=0$) in $\CC^n$.
Similarly, if the hypersurface is in an orbifold
$\CC^{n-1}/\Gm \times\CC^2_{uv}$, the same T-duality%
\ftl{f-T}{In both cases, $X^{2n}$ can be viewed as a fibration $uv=$const.,
the base being $\CC^{n-1}$ or $\CC^{n-1}/\Gm$, respectively
(for $uv\neq0$, the fiber is $\CC^*$).
T-duality in the fiber leads to the above NS5 configurations.}
will give an NS5 brane
wrapped on $\Sg$ ($H=0$) in $\CC^{n-1}/\Gm \times\CC$.
In both cases, the singularity of $X^{2n}$ leads to a corresponding
singularity of $\Sg$.
As argued above, before the T-duality, the non-trivial part of the
decoupled dynamics is confined to the singular region of $X^{2n}$.
For the T-dual NS5 brane configuration, this means that its decoupled
dynamics is localized in the neighborhood of the singularity at $\Sg$,
in agreement with the general expectations.
The D branes wrapped on vanishing cycles in $X^{2n}$ are dual to D-branes
that end (have boundaries) on the NS5 brane in the singular region
\cite{KLMVW9604}.

\newsection{A Singularity on a Hypersurface in an Orbifold}
\secl{s-orb}%
In this section, we consider the holographic duality (described in the
introduction) for singularities modeled by a hypersurface in an
orbifold. The holographicly-dual pair is identified in the first subsection.
Then, properties of the decoupled theory, as obtained from both dual
descriptions, are discussed,
first for a general dimension $d$ (in subsection  \ref{s-check}) and then for
$d=4$ (in subsection \ref{s-four}). This serves both to check the proposed
duality and as a preparation for the analysis in section \ref{s-SGT}.

\subsection{The Holographicly-Dual Pair}
\secl{s-pair}%
We start by reviewing the case with a hypersurface in flat space,
considered in \cite{GKP9907}\cite{GK9909}\cite{GK9911}, 
and then introduce the orbifold.

\newpar{A Hypersurface in Flat Space}
\nl A large class of singular manifolds $X^{2n}$ can be modeled by a
hypersurface
\[ W(z_1,\ldots,z_{n+1})=0 \]
in $\CC^{n+1}$, with $W$ a quasi-homogeneous polynomial,
\ie, transforming as $W\goto\lm W$ under
\beql{Ctrans} z_a\goto\lm^{r_a}z_a \hsc \lm\in\CC \hs, \eeq
for some positive (rational) ``weights'' $r_a$.
The authors of \cite{GKP9907} considered such a singularity which has the
following additional properties:
\begin{itemize}
\item An isolated singularity (point-like; at $z=0$);
the condition for this is that the polynomial $W$ is ``transverse''
\beql{trans} dW=0 \hs\hs\Leftrightarrow\hs\hs z=0 \hs. \eeq
\item A singularity that appears at finite distance in the moduli space of
(compact, smooth) CY manifolds; as shown in \cite{GVW9906}, the condition for
this is $r_\Om>0$, where
\beql{r-Om} r_\Om=\sum_a r_a-1 \hs. \eeq
\end{itemize}

It was proposed in \cite{GKP9907} that the dynamics at the singularity in the
decoupling limit is described by type II string theory on
\beql{vac-f} \RR^{d-1,1}\times \RR_\phi \times U(1)_Y \times LG_W \hs, \eeq
where $\RR_\phi \times U(1)_Y$ is the $N=2$ Liouville theory and $LG_W$ is the
$N=2$ Landau-Ginzburg (LG) model with the superpotential being the polynomial
$W$. One way to understand this identification is as follows:
to obtain better control on the model, one deforms $X^{2n}$ to a smooth
hypersurface $X^{2n}_\mu$ by changing $W$
\beql{Wmu} W\goto W_\mu=W+\mu V \eeq
(where $V$ is an appropriate quasi-homogeneous polynomial%
\footnote{We assume quasi-homogeneity for simplicity.
The arguments below generalize in a straight-forward way to a general
polynomial, which is of the form $V=\sum_i V_i$, where $V_i$ are
quasi-homogeneous. Note that the $z_0$-dressing in eq. (\ref{hWmu}) will not
be the same for all terms.},
with weight $r_V$).
Next, this deformed hypersurface is embedded in a weighted projective space:
$\CC^{n+1}$ is embedded in $\WCP^{n+1}_{-r_\Om, r_1,\ldots,r_{n+1}}$,
by introducing an additional coordinate $z_0$, with a {\em negative} weight
$-r_\Om$ (where $r_\Om$ is defined in eq. (\ref{r-Om}));
\internote{The weight $r_0=-r_\Om$ is determined by the CY condition
$r_0+\sum r_a =r_F=1$. Why is this relevant?}
to embed $X^{2n}_\mu$ in this space, $W_\mu$ is made quasi-homogeneous,
by an appropriate $z_0$ ``dressing''%
\footnote{Note that, to deform the singularity, $V$ must dominate at $z\goto0$.
This means that $r_V<1$, so the power of $z_0$ in (\ref{hWmu}) is
{\em negative}!}:
\beql{hWmu} W_\mu\goto\hat{W}_\mu=W+\mu\hat{V} \hsc
\hat{V}=z_0^{(r_V-1)/r_\Om}V \hs. \eeq
Now one argues, following \cite{GV9506}\cite{OV9511},
that the SUSY NL$\sg$M on $X^{2n}_\mu$ is described,
in the IR limit, by a  LG model with the superpotential $\hat{W}_\mu$
(eq. (\ref{hWmu}))%
\footnote{More precisely, this is an orbifold of the above LG model,
with a projection on integral $U(1)$ charges of the N=2 superconformal
algebra. However, in the present context (of string theory on (\ref{vac-f})),
this orbifold is part of the GSO projection, so it can be ignored at this
stage.}
\footnote{In \cite{GKP9907}, this identification was obtained, for $\mu=0$,
using an embedding in a gauged linear $\sg$ model, following \cite{Witten93}.}.
\internote{
\nl Probably, the exact statement is that the LG is the IR limit of the CY.
\nlb This is suggested by the approach of \cite{GKP9907}.
\nlb Can this be proven, using the gauged linear $\sg$ model approach?}
Finally, one identifies the $z_0$ CFT with $N=2$ Liouville theory%
\footnote{Strictly speaking, this is Liouville theory when $V=1$,
in which case, $\dl\hat{W}=\mu\hat{V}=\mu e^{-\hat{\phi}/Q}$ is the Liouville
interaction.
This is the case considered in \cite{GV9506}\cite{OV9511}, and the $z_0$
CFT was identified there with the $SL(2)/U(1)$ coset SCFT
(at level $k=1/r_\Om$;
see \cite{OV9511} for a review of the evidence for this identification).
On the other hand, the analysis in \cite{GKP9907} suggested the identification
with Liouville theory, as described above. This led to a proposal
\cite{GK9909} that these two theories are equivalent.}.
This theory has a single chiral superfield $\hat{\phi}$.
Denoting the (scalar) bottom component of $\hat{\phi}$ by $\phi+iY$,
the dilaton $\Phi$ is proportional to $\phi$:
\beql{Phi} \Phi=-\frac{Q}{2}\phi \hs, \eeq
where $Q$ is a positive parameter.
Now, the identification is
\beql{zQ} z_0=e^{\frac{Q}{2} \hat{\phi}} \hsc Q=\sqrt{2r_\Om} \hs. \eeq
\internote{Details:
\nlb $Q\leftrightarrow r_\Om$: from comparing central charges:
\nl $3(1+Q^2)=3(1-2r_0)=3(1+2r_\Om)$.
\nlb $\hat{\phi}\leftrightarrow z_0$: from comparing conformal dimensions
\nl (\eg, $\Dl=\half$ for the dressing of a constant).} 
Recall that $r_\Om$ was chosen to be positive (for the singularity to be at
finite distance in the CY moduli space), and this is indeed required for the
consistency of the above identification.
Thus, combining these relations, one obtains the background (\ref{vac-f}).

After obtaining the above identification, one may want to remove the
deformation $\mu V$, returning to the original, singular, hypersurface
$X^{2n}$.
However, one finds that the worldsheet formulation of the string theory
in the background (\ref{vac-f}) becomes singular.
Indeed, the string coupling $g_s$ is related to the dilaton: $g_s=e^\Phi$,
so eq. (\ref{Phi}) implies that $g_s$ diverges at $\phi\goto-\infty$.
For $\mu=0$, the superpotential is independent of $\phi$, so the region
$\phi\goto-\infty$ is accessible and the perturbative expansion is singular.
This singularity is a reflection \cite{Witten9507}\cite{OV9511}
of the existence of massless solitons:
D-branes wrapped on the vanishing cycles at the singularity \cite{Strom9504}.
Therefore, to obtain a well-behaved perturbative description, it is necessary
to make these solitons massive.

This is achieved by the double scaling limit (\ref{dbl-lim}), suggested in
\cite{GK9909}\cite{GK9911}: the deformation discussed there
was with $V=1$ in eq. (\ref{Wmu}) and the quantity held fixed was the ratio
\beql{m-def} m=\frac{\mu^{r_\Om}}{g_s} \eeq
($r=r_\Om$ in eq. (\ref{dbl-lim})).
To understand the meaning of $m$, one notes that $X^{2n}_\mu$ has
holomorphic $n$-cycles that vanish for $\mu=0$.
The volume $V_C$ of such a cycle $C$ is $V_C=\int_C\Om$, where $\Om$ is the
holomorphic $n$-form  on $X^{2n}_\mu$. 
{}From the explicit expression for $\Om$
\beql{Om} \Om=\frac{dz_1\wedge\ldots\wedge dz_{n+1}}{dW} \hs, \eeq
one finds that the weight of $\Om$ under (\ref{Ctrans}) is $r_\Om$, defined
in eq. (\ref{r-Om}) (and assumed to be positive). This implies that
$V_C\sim\mu^{r_\Om}$ and, therefore, the $d$-dimensional mass/tension of a
D-brane wrapped on $C$ is $V_C/g_s\sim m$.
Thus, in the double scaling limit with $m\neq0$,
there are no massless solitons,
so one can expect that the worldsheet formulation will be under control
and indeed it is:
the strong coupling region $\phi\goto-\infty$ is inaccessible,
because of the deformed superpotential $\hat{W}$ that diverges there.
Using the relation $g_s=e^\Phi$, one can show that the theory depends on
$g_s$ and $\mu$ only through the ratio $m$ defined in eq. (\ref{m-def})
and it can be studied perturbatively, $1/m$ being the string loop expansion
parameter.

\newpar{The Orbifold}
\nl Considering the above scenario,
let $\Gm$ be a finite subgroup of $SU(n+1)$ which commutes with (\ref{Ctrans})
(\ie, mixes only coordinates with the same weight)
and leaves the polynomial $W$ invariant.
The equation $W=0$ is, therefore, well defined in the orbifold
$\CC^{n+1}/\Gm$, and it defines there a hypersurface $X^{2n}$, with an
isolated singularity at $z=0$. We are interested in the $d$-dimensional
theory describing the decoupled dynamics near this singularity.
Note that the singular point $z=0$ of $X^{2n}$ is at the orbifold singularity,
which is what is needed for the decoupled theory to be affected by the
orbifold.
To obtain an anti-holographic formulation of this theory, we replace the
$LG_W$ factor in the background (\ref{vac-f}) by the orbifold $LG_W/\Gm$.
Note that $\Gm$ defines a symmetry group of $LG_W$ (acting trivially on
the supercharges), so one can orbifold by it.
We, therefore, propose (generalizing \cite{GKP9907}) that the following
theories are two formulations of {\em the same} $d$-dimensional theory:
\begin{itemize}
\item {\em geometric formulation}: string theory on
\beql{vac-geo} \RR^{d-1,1}\times X^{2n} \hs, \eeq
in the decoupling limit (as described in section \ref{s-dec});
\item {\em worldsheet formulation}: string theory on 
\beql{vac-o}
\RR^{d-1,1}\times \RR_\phi \times U(1)_Y \times LG_W/\Gm \hs.
\eeq
\end{itemize}

\subsection{Aspects of the Holographic Relation}
\secl{s-check}
In this and the next subsections, properties of the decoupled $d$-dimensional
theory are identified in the worldsheet formulation.
This will be used, in the next section, to analyze specific four-dimensional
theories.
Some aspects of this identification depend only on the
$\RR^{d-1,1}\times \RR_\phi \times U(1)_Y$ factor of the background
(\ref{vac-o}), while the factor $LG_W/\Gm$ could be replaced by any 2D (2,2)
SCFT. These aspects are, therefore, naturally the same here as in the
unorbifolded case, discussed in \cite{GKP9907}.
Other aspects depend also on properties of the $LG_W/\Gm$ factor and,
accordingly, on the correct identification of the $d$ dimensional theory.
As in \cite{GKP9907}, these aspects are compared to the information available
from the geometric formulation and agreement is found, providing evidence
for the duality.
The results and their implications are summarized in section \ref{s-sum}.

\newpar{Space-Time Supersymmetry}
\nl We start with supersymmetry, showing that in both formulations there are
$2^{\frac{d}{2}+1}$ preserved space-time supercharges.

In the worldsheet formulation, the orbifold action in $LG_W$ was chosen to
commute with the supercharges and the transformation (\ref{Ctrans}) so,
in the IR CFT, it commutes with the full $(2,2)$ superconformal algebra
(in which one of the $U(1)$ R-symmetries is identified with (\ref{Ctrans})).
Therefore, the CFT (\ref{vac-o}) has $(2,2)$ supersymmetry and, as in any
such situation, one can construct the space-time supercharges using the
$U(1)$ currents of the superconformal algebra.
This leads, as before orbifolding (see \cite{GKP9907} for more details),
to $2^{\frac{d}{2}+1}$ space-time supercharges.
The unimodularity of the elements of $\Gm$ (det($g$)=1, $\forall g\in\Gm$)
insures that, for an appropriate choice of the orbifold action%
\ftl{Kchoice}{See \cite{IV90} for more details
(in the notation there, the choice is $(-1)^{K_g}=\det(g)$).},
the space-time supercharges are $\Gm$-invariant.

In the geometric formulation, before orbifolding there were
$2^{\frac{d}{2}+1}$ supercharges (as in any CY compactification),
so it remains to check the effect of the orbifold.
The supercharges are related to the
holomorphic $n$-form $\Om$ on $X^{2n}$:
indeed, $\Om$ can be written in terms of a covariantly constant spinor
$\eta$ on $X^{2n}$ as $\Om_{i_1\ldots i_n}=\eta^t \Gm_{i_1\ldots i_n}\eta$,
so the orbifold actions on $\Om$ and on the supercharges are related.
Using the explicit expression (\ref{Om}) for $\Om$, one finds that $\Om$ is
$\Gm$-invariant: both numerator and denominator are invariant
(the first invariance follows from the unimodularity of the elements of $\Gm$).
Therefore, the orbifold can have at most a $\ZZ_2$ action on the supercharges.
This $\ZZ_2$-ambiguity has the following meaning: the supercharges appear
quadratically in the supersymmetry algebra, so with a given $\Gm$-action
on the bosonic coordinates, its action on the supercharges
always has a $\ZZ_2$ ambiguity.
The invariance of $\Om$ means that one can choose the $\Gm$-action on
the supercharges to be trivial and, with this choice, the orbifold
does not break supersymmetry%
\footnote{This choice is the space-time analog of the choice made in the
worldsheet formulation (see footnote \ref{Kchoice}).}.

\newpar{Deformations Preserving the Space-Time Supersymmetry}
\nl Next we discuss SUSY-preserving deformations.

Consider, for example, a deformation of the polynomial $W$: $\dl W=\mu V$
(where $V$ is a quasi homogeneous polynomial, with weight $r_V$; as in
eq. (\ref{Wmu})).
In the geometric formulation, it induces a change in the manifold $X^{2n}$
and, therefore, also in the corresponding decoupled theory.
More specifically, the coefficient (modulus) $\mu$ parametrizes a change in
the complex structure of $X^{2n}$, a deformation that does not break
supersymmetry.
In the worldsheet formulation, a change in $W$ induces a change in the
superpotential: $\dl\hat{W}=\mu\hat{V}$, as in eq. (\ref{hWmu}).
This means that one adds to the worldsheet Lagrangian a top component of
a chiral-chiral superfield (of the worldsheet $N=2$ SCFT).
The $z_0$ dressing insures that this operator has dimension
$\Dl=\tilde{\Dl}=1$.
Such a change in the Lagrangian preserves the worldsheet superconformal
symmetry \cite{Dixon87} and, therefore,
defines a (truly) marginal perturbation, parametrized by the coefficient
(coupling) $\mu$, which preserves the space-time supersymmetry.

This can be generalized for any chiral-chiral primary operator $V$ in the
$LG_W/\Gm$ factor with equal left and right $U(1)$ charges $f_V=\tilde{f}_V$.
$V$ can be ``dressed'' by the Liouville fields:
\beql{chiral} \hat{V}=e^{\bt(\phi+iY)}V \hs. \eeq
This is a chiral-chiral primary of the worldsheet $N=2$ SCFT, which is the
bottom component of a chiral-chiral superfield
(we denote both by the same symbol $\hat{V}$).
The $U(1)$ charge of $\hat{V}$ (both left and right) is
$f_V-\bt Q$ and, to obtain a marginal deformation, as described above,
one chooses $\bt$ to set this charge to 1:
\beql{qfV} \bt=\frac{f_V-1}{Q} \hs. \eeq
This defines a map from the ring of chiral-chiral primary operators
(the (c,c) ring) in the $LG_W/\Gm$ factor, to SUSY-preserving changes in
the decoupled theory.

Before orbifolding, the (c,c) ring in $LG_W$ is spanned by quasi homogeneous
polynomials $V(z_a)$, identified modulo $\pt_a W$.
\internote{By the equations of motion,
$\pt_a W\sim D\dg\bar{D}\dg z_a\dg$,
which is a descendent (non-primary) chiral [Lerche,Vafa,Warner89].}
Their $U(1)$ charges $f_V,\tilde{f}_V$ are
equal and coincide with the weight $r_V$ of $V$ under (\ref{Ctrans}).
In the geometric formulation, the corresponding deformation was identified in
\cite{GKP9907} with a deformation of the complex structure in $X^{2n}$,
as described above.
\internote{
For this to be a bijective correspondence, the deformations of
$W$ should also be identified modulo $\pt_a W$.
\nls Adding $dW$ is indeed trivial (generates a translation),
but why is $VdW$ trivial for a non-constant $V$?
\nls $\dl W\sim V\pt_a W$ can be undone by a diffeomorphism $\dl z=V$
but this is not, in general, a symmetry of $\CC^{n+1}$,
so it is not clear why this should be moded out.
\nlb The resulting ring $\Rc=\CC[z_a]/dW$ is called ``the singularity ring''.
\nls Possible reference: V.I. Arnold, Gusein-Zade and Varchenko,
``Singularities of Differential Maps'', Boston: Birkh\"{a}user, 1988.}%
The orbifold projection truncates this set of deformations in the same
way in both the worldsheet and geometric formulations, by restricting the
polynomials $V$ to be $\Gm$-invariant.
This can be viewed as further evidence for both the holographic relation and
the geometric identification of these deformations.
Additional (c,c) operators appear in twisted sectors of the orbifold.
The corresponding deformations in the $d$-dimensional theory are related
to its influence by the orbifold singularity.
This will be seen in specific examples in the next section.

\newpar{Couplings vs. Moduli}
\nl In the previous paragraph, possible deformations of the
$d$-dimensional theory were discussed, as viewed in its two formulations.
The parameter $\mu$ parametrizing such a deformation can be either
\begin{itemize}
\item a coupling: parametrizing a change in the theory (\eg, a coefficient
of a term in the action); or
\item a modulus: parametrizing a change in the vacuum (\eg, a vacuum
expectation value (vev)).
\end{itemize}

In the holographic duality with $AdS$ vacua
(for a review, see \cite{AGMOO9905}),
considered in the framework of semi-classical supergravity,
the distinction between couplings and moduli is related to the behavior of
the corresponding bulk fields near the boundary:
non normalizable modes are related to couplings in the boundary theory
\cite{GKP9802}\cite{Witten9802} while normalizable modes are related to
vevs in the boundary \cite{BKL9805}\cite{BKLT9808} (see also \cite{KW9905}).
It is natural to expect the same distinction in the present holographic
duality%
\footnote{The relation to $AdS_{d+1}$ vacua can be made more concrete in
the $d=2$ case, by using the correspondence, described in \cite{GKP9907},
between $AdS_3$ vacua and $\RR^{1,1}\times\RR_\phi$ vacua.}.
Here, the ``boundary'' is at $\phi\goto\infty$ and, for a deformation
defined by a vertex operator $\hat{V}$, the deformation parameter $\mu$
is identified as a coupling iff the corresponding wave function is
non-normalizable at $\phi\goto\infty$.
For the vertex operators (\ref{chiral}), this implies that for
\beql{btQ} \bt>-\frac{Q}{2} \hs, \eeq
$\mu$ is a coupling and for $\bt<-\frac{Q}{2}$, it is a modulus%
\footnote{To obtain these conditions, one notes that the vertex operator
$\hat{V}$ is related to the wavefunction $\Psi$ by $\hat{V}=g_s\Psi$, where
$g_s=e^\Phi=e^{-\frac{Q}{2}\phi}$.}.
The case $\bt=-\frac{Q}{2}$ is more delicate and will be discussed below.
The bound (\ref{btQ}) appears also when the string theory on (\ref{vac-o})
is viewed as a 2D CFT coupled to quantum gravity. In that context, (\ref{btQ})
is the condition for the operator (\ref{chiral}) to exist as a local operator
in the theory (see \cite{Seiberg90} for more details).

Using eqs. (\ref{qfV}),(\ref{zQ}), one obtains that for
\beql{coupl} f_V>1-r_\Om \hs, \eeq
$\mu$ is a coupling and for
\beql{modul} f_V<1-r_\Om \hs, \eeq
it is a modulus.
This applies, in particular, to the deformations of the polynomial $W$:
$\dl W=\mu V$.
For these deformations, the distinction between couplings and moduli
was investigated in the geometric formulation in \cite{GVW9906}
(see also \cite{SV9910}). In this approach, a coupling is a 10D mode whose
$d$-dimensional kinetic energy diverges and, consequently, its fluctuations
are ``frozen''.
For $r_V\neq1-r_\Om$, this criterion led to the same conditions on $r_V$ as
above, namely, the conditions (\ref{coupl}),(\ref{modul}) with $f_V=r_V$.
The deformation with $r_V=1-r_\Om$ is found to be a coupling.
It is interesting to note that also in this approach,
$\mu$ was identified as a non-fluctuating coupling when the
perturbation was supported (in some sense) far from the singularity
(which is indeed identified here with $\phi\goto\infty$).
\internote{Details: [Shapere\&Vafa9910]
\nlb Defining the $n$-form $\Om_i=\pt\Om/\pt\mu_i$
($r_{\Om_i}=r_\Om-r_{\mu_i}=r_{V_i}+r_\Om-1$),
\nl $\mu$ is a vev
$\Leftrightarrow$ (why?) $\int|\Om_i|^2$ is normalizable at large $z$
$\Leftrightarrow$ $r_{\Om_i}<0$.
\nlb For $r_\Om>0$ this means that for all vevs, $r_V<1$
\nl (corresponding to a deformation that dominates over $W$ at $z\goto0$).
\nlb sp$\{\Om_i\}$ contains {\em all} the normalizable $n$-forms localized
near the singularity.}
The manifold considered in \cite{GVW9906} was a hypersurface in
{\em flat space}.
However, the aspects analyzed were convergence of certain integrals,
and these aspects are not influenced by a finite group of identifications.
Therefore, the results of \cite{GVW9906} apply also to the present case.

\newpar{R-Symmetry}
\nl In the worldsheet formulation, the scalar $Y$ is free, and this leads to
two $U(1)$ symmetries, with the following conserved charges
\beql{Rsym}
R=i\frac{2}{Q}\oint\pt Y \hsc \tilde{R}=i\frac{2}{Q}\oint\bar{\pt} Y \hs.
\eeq
The space-time supercharges are charged under these symmetries,
(see \cite{GKP9907} for details), so these are R-symmetries.
Defining the linear combinations
\beql{Rpm} R_{\pm}=R\pm\tilde{R} \hs, \eeq
all supercharges have charge $|R_{\pm}|=1$.
For the deformations defined by $\hat{V}$ in eq. (\ref{chiral}),
$R=\tilde{R}$, so
\beql{R-ws}  R_-(\mu)=0 \hsc
R_+(\mu)=- R_+(\hat{V})=-2R(\hat{V})=-\frac{4\bt}{Q}
=2\frac{1-f_V}{r_\Om} \eeq
(using eqs. (\ref{chiral}),(\ref{qfV}),(\ref{zQ}) and the fact that the
worldsheet action is invariant).
Combining eqs. (\ref{coupl}),(\ref{modul}),(\ref{R-ws}), one obtains that
$\mu$ is a coupling when $R_+(\mu)<2$ and a modulus when $R_+(\mu)>2$ .

In the geometric formulation, the transformation (\ref{Ctrans}) with $|\lm|=1$
is an isometry of $X^{2n}$ and, therefore, induces a $U(1)$ symmetry in the
corresponding theory. To identify its action on supercharges, one considers
again the holomorphic $n$-form $\Om$. {}From eq. (\ref{Om}) one finds that
the weight of $\Om$ under (\ref{Ctrans}) is $r_\Om$, defined in eq.
(\ref{r-Om}), and the fact that it is non-zero means that the above $U(1)$
symmetry is an R-symmetry. With a normalization in which the supercharges
have $U(1)$ charge $\pm1$, the R-charge $ R'_+$ is related to the weight $r$
by
\[R'_+=2\frac{r}{r_\Om} \hs. \]

It was suggested in \cite{GKP9907} that $R'_+$ should be identified with $R_+$.
This was checked by verifying that, for deformations of $W$: $\dl W=\mu V$
(where $f_V=r_V$), the geometric formulation indeed gives the same R-charge as
obtained in eq. (\ref{R-ws}):
\beql{R-geo}  R'_+(\mu)=2\frac{r_\mu}{r_\Om}=2\frac{1-r_V}{r_\Om} \eeq
(where the last equality follows from the fact that the polynomial $W$
has weight $r_W=1$).

\subsection{Four-Dimensional Theories}
\secl{s-four}%
We move now to a closer look at the case $d=4$.
The number of supercharges is $8$, corresponding to
$N=2$ supersymmetry in four dimensions.

\newpar{Coupling-Moduli Pairing}
\nl The (c,c) ring of the $LG_W/\Gm$ CFT has a $\ZZ_2$
``reflection'' symmetry, relating operators with $U(1)$ charges $f$ and
$\hat{c}-f$, where $3\hat{c}$ is the central charge of the CFT:
\beql{hc} \hat{c}=\sum_a(1-2r_a)=(n-1)-2r_\Om \hs. \eeq
To identify this symmetry, one uses two bijective relations between the (c,c)
ring and the (a,a) ring -- the ring of antichiral-antichiral primary
operators. One relation is charge conjugation
\internote{Details:
\nlb Charge conjugation (reversing all $U(1)$ charges) exists, by definition,
in any CFT with a non-degenerate Zamolodchikov metric.
\nlb Under charge conjugation: $(f,\tilde{f})\goto(-f,-\tilde{f})$.
\nlb In a unitary CFT, this means $R\leftrightarrow R$, $c\leftrightarrow a$.}
and the other is obtained using the spectral flow of the $N=2$ superconformal
algebra (the existence of this second relation is a consequence of the
unimodularity of the elements of $\Gm$ \cite{IV90}).
Combining these two, one obtains a bijection in the (c,c) ring.
As to the $U(1)$ charge, starting with a (c,c) operator with charge $f$,
complex conjugation gives an (a,a) operator with charge $-f$ and then the
spectral flow, gives a (c,c) operator with charge $\hat{c}-f$.
\internote{Comments for $\Gm=1$:
\nlb The reflection symmetry is manifested in the 
generating function (Poincare polynomial)
$\tr_{(c,c)}t^r=\prod_a\frac{1-t^{1-r_a}}{1-t^{r_a}}$
\nlb This function also gives the dimension of the (c,c) ring:
$\prod_i\frac{1-r_a}{r_a}$
\nl (by taking the $t\goto1$ limit).}

This reflection symmetry is not specific to models related to
four-dimensional singularities. However, its significance is enhanced for
$d=4$, as we now explain.
In this case, $\hat{c}=2(1-r_\Om)$ (using eq. (\ref{hc}) with $n=3$).
Comparing to the bounds (\ref{coupl}),(\ref{modul}), one can see that,
for operators with $f\neq\hat{c}/2$, the reflection symmetry induces
a pairing between coupling deformations and moduli deformations.
This pairing has a natural interpretation in the 4D theory
(as was also observed independently in \cite{SV9910}).
The deformations defined by (c,c) operators are related, in the 4D theory
to scalars in vector superfields.
\internote{
\nlb This is known to be true for complex deformations in
any type IIB string theory on $\RR^{3,1}\times CY_3$.
\nlb This follows also from the relation to SW theory, where all parameters
in $H$ -- masses and vevs -- are related to such scalars.}%
When the deformation parameter $\mu$ is a coupling with $R_+(\mu)<2$,
it corresponds to adding a {\em top component} $\Ac_t$ of a vector
superfield $\Ac$ to the prepotential (with $\mu$ as a coefficient),
while if it is a modulus with $R_+(\mu)>2$, the change is in the vev
of a {\em bottom component} $\Ac_b$ of a vector superfield $\Ac$.
\internote{
\nlb This follows from preservation of supersymmetry
\nlb This is true also for $d\neq4$, but only for $d=4$ these two
types of deformations can be paired.}
Thus, each vector superfield $\Ac$ in the 4D theory defines
two deformations - one coupling and one modulus -- and it is natural to
identify these pairs with the pairs seen in the worldsheet formulation.

Evidence for this identification is obtained by considering R-charges.
The 4D $N=2$ supersymmetry algebra has a $U(1)\times SU(2)$ R symmetry
group.
In the previous subsections, a $U(1)_+\times U(1)_-$ R-symmetry group was
found (with charges (\ref{Rsym}).
To identify the relation between these two groups, one notes that the
scalars in vector superfields are neutral under the $SU(2)$ factor of
the R-symmetry and charged under the $U(1)$ factor.
The charges found in subsection \ref{s-check} (eq. (\ref{R-ws}))
imply that $R_+$ should be identified with the $U(1)$ factor and $R_-$ with
a $U(1)$ subgroup of the $SU(2)$ factor.
Now, considering a pair $\mu_c,\mu_m$ of deformation parameters, related by
the reflection symmetry described above, one finds in both approaches that
the sum of their R-charges is four, providing evidence for their
correspondence to the same vector superfield $\Ac$.
In the worldsheet formulation this follows from  eq. (\ref{R-ws}),
while in the 4D field theory this follows from%
\footnote{The $U(1)$ factor of the R-symmetry is defined only up to a shift
by a $U(1)$ symmetry that commutes with the supercharges.
However, the relations (\ref{Rmumc}) are invariant under such a shift,
so they indeed provide evidence for the proposed interpretation of the pair
of deformations.}:
\beql{Rmumc}
R_+(\mu_m)=R_+(\Ac_b) \hsc R_+(\mu_c)=-R_+(\Ac_t)=-[R_+(\Ac_b)-4] \hs.
\eeq

\newpar{IR Conformal Dimensions and Unitarity}

In the extreme IR limit, one obtains an $N=2$ superconformal field theory.
The 4D $N=2$ superconformal algebra (SCA) includes a $U(1)\times SU(2)$
R-symmetry and implies a relation between the R-symmetry quantum numbers of
an operator and its conformal dimension \cite{DP85}.
In particular, the bottom component of a vector superfield is a chiral
primary field of the SCA, a scalar of the $SU(2)$ R-symmetry and a Lorentz
scalar.
For such a field, the conformal dimension is $D=\half R''_+$,
where $ R''_+$ is the charge of the $U(1)$ R-symmetry.
\internote{Details: in a general 4D $N=2$ SCFT \cite{APSW9511}
\nlb Fields are labeled by the Lorentz spin $(j,\tilde{j})$,
the conformal dimension $D$ the $U(1)_R$ charge $ R_+$
and the $SU(2)_R$ spin $I$.
\nlb A chiral field is characterized by $j=0$
\nlb For a chiral primary field: $D=2I+\half R_+$; $\half R_+\ge j+1$.}
Identifying $R''_+$ with the charge $R_+$ (\ref{Rpm})%
\footnote{This identification will be discussed further at the end of this
section.},
one can determine the conformal dimensions of the 4D vector superfield
related to a deformation defined by a worldsheet (c,c) operator $V$:
if $\mu=\mu_c$ is a coupling with $R_+(\mu)<2$,
\beql{dim-c}
D(\Ac_b)=\frac{1}{2} R_+(\Ac_b)=\frac{1}{2}[R_+(\Ac_t)+4]
=\frac{1}{2}[- R_+(\mu_c)+4]=\frac{f_V-1}{r_\Om}+2
\eeq
and if $\mu=\mu_m$ is a modulus with $R_+(\mu)>2$,
\beql{dim-m}
D(\Ac_b)=\frac{1}{2} R_+(\Ac_b)=\frac{1}{2} R_+(\mu_m)=\frac{1-f_V}{r_\Om}
\hs, \eeq
(where, in the last stage, we used (\ref{R-ws})). 

As observed in \cite{GKP9907}, the bound (\ref{coupl}) for deformations that
are couplings gives, when substituted in (\ref{dim-c})),
$D(\Ac_b)>1$, which is (almost; see below) the unitarity bound on the
conformal dimension of $\Ac_b$.
The same is true for deformations that are moduli: the bound is (\ref{modul})
and when it is substituted in (\ref{dim-m})), it also gives%
\footnote{This was observed independently also in \cite{SV9910}.}
$D(\Ac_b)>1$.
\internote{
\nlb For couplings, the same relation can be derived in 2D with (2,2)
space-time SUSY.
\nls There, the identification of R-charges is also implied by the
relation to $AdS_3$, where these charges are explicitly constructed.
\nls Does such a relation exist for $d>2$?
\nlb In the remaining cases there is no $U(1)$ factor but, presumably,
also there a unitarity bound is relevant.
\nlb For vevs, a generalization to $d\neq4$ was suggested in \cite{SV9910},
but it seems wrong.}

\newpar{Deformations With $R_+(\mu)=2$}
\nl So far, we only discussed the identification, in the 4D theory,
of deformations with $R_+(\mu)\neq2$. We turn now to those with $R_+(\mu)=2$.
The precise unitarity bound is $D(\Ac_b)\ge1$ so, for $R_+(\mu)=2$, it allows
both possibilities -- a coupling and a modulus.
In either case, this leads to the identification of a vector superfield
$\Ac$ with $D(\Ac_b)=1$. The SCA implies that such a superfield is free.
The geometric formulation suggests \cite{GVW9906} that $\mu$ is a coupling
(as described in the previous subsection)
so its natural interpretation would be as that for $R_+(\mu)<2$:
a coefficient of a term $\Ac$ in the superpotential.
However, as observed in \cite{APSW9511} (using $D(\Ac)=1$ and the SCA),
the corresponding contribution to the  Lagrangian is a total derivative and,
therefore, has no effect.
\internote{Further information \cite{APSW9511}:
\nl Expanding the prepotential $\Fc\sim\mu\Bc^j$ in the free vector
superfield $\Bc$,
\nl the possibilities are (after dropping total derivatives):
\nls $j=2$ $\goto$ $\mu\sim\dl\tau$ (the gauge coupling; marginal)
\nls $j>2$ $\goto$ $D(\mu)<0$ (irrelevant).}%
Instead, we propose, following \cite{APSW9511}, that in this case,
$\mu$ corresponds to a vev (of a bottom component) of a vector superfield
$\Ac$ with the following properties:
\begin{itemize}
\item $\Ac$ is {\em frozen} in the IR limit, \ie,
its kinetic energy diverges in this limit;
\item $\Ac$ couples to a conserved current which is non-trivial
in the IR SCFT, \ie, the corresponding symmetry, which becomes a global
symmetry in the IR, acts non-trivially in this SCFT.
\end{itemize}
The simplest realization of such a situation is $N=2$ SQED: a single
vector multiplet (leading to a $U(1)$ gauge theory) with a charged massless
hypermultiplet. Because of the non-trivial charge, the gauge coupling
vanishes in the IR limit, leading to a divergent kinetic energy.

According to the above proposal, $\mu$ is a coupling in the IR SCFT,
in agreement with the analysis in the geometric formulation.
However, it originates from a modulus.
Moreover, it may become a modulus also in the IR, upon deformation:
for example, in SQED, deforming the theory by giving a vev $\mu$ to
(the bottom component of) the vector superfield leads to a massive
hypermultiplet. In the deformed theory, the gauge coupling does not vanish in
the IR and, consequently $\mu$ remains a vev of a fluctuating (although free)
field.
This mixed nature of $\mu$ is, in fact, suggested also by the $\phi$ dressing
in the worldsheet formulation, as described in the previous subsection.
For $\bt>-\frac{Q}{2}$, the
wave function is exponentially supported at $\phi\goto\infty$ and vanishes
at $\phi\goto-\infty$, while for $\bt<-\frac{Q}{2}$, the situation is reversed.
The present case corresponds to $\bt=-\frac{Q}{2}$ and it is special in having
support at both regions.

The above identification also leads to the correct conformal dimension for
$\mu$. This can be argued as follows \cite{APSW9511}: since a conserved charge
is dimensionless, the corresponding conserved current has dimension three,
so the vector field coupling to it has dimension one; the SCA now implies that
the bottom component of the corresponding superfield has also dimension one.
\internote{Details:
$D(\int j_\mu)=0\hs\Goto D(j_\mu)=3$
\nl $\Goto D(A_\mu)=1\hs\Goto D(F_{\mu\nu})=2\hs\Goto D(\Ac_b)=1$.}
We end with some comments:
\begin{itemize}
\item
The dimension of the parameter $\mu$ is, by definition, $D(\mu)=D(\Ac_b)$
if it is a modulus; and $D(\mu)=4-D(\Ac_t)$ if it is a coupling. In both
cases, this gives
\beql{DR} D(\mu)=\frac{1}{2}R_+(\mu) \hs, \eeq
therefore,
\begin{center}
$\mu$ is a coupling $\hs\hs\Longleftrightarrow\hs\hs D(\mu)\le1$ .
\end{center}
\internote{Details:
\nl $D(\mu_c)=4-D(\Ac_t)=4-[D(\Ac_b)+2]=\half[4-R_+(\Ac_b)]=-\half R_+(\Ac_t)
=\half R_+(\mu_c)$}
\item
As explained above, a free vector superfield $\Ac$
(with $D(\Ac)=1$) leads only to a single deformation of the theory. In the
worldsheet formulation this means that one should not expect to identify,
in the $LG_W/\Gm$ CFT, a pairing between (c,c) operators with charge
$f=\hat{c}$. Indeed, one finds that the number of such operators is not
always even.
\item
It was suggested in \cite{APSW9511} that a coupling $\mu$ with $D(\mu)<1$
can also be identified with a vev of a vector superfield $\Bc$ that, in the
IR limit, is frozen but dues not couple to a non-trivial conserved current.
Such a superfield would couple to the SCFT through a term $\Ac\Bc/\Lm^\dl$
in the prepotential, where $\Ac$ is an interacting vector superfield in
the SCFT and $\Lm$ is some scale in the underlying theory
(note that $\dl=D(\Ac)-1>0$, so this is an irrelevant interaction).
The parameter $\mu$ is then identified as $\mu=\vev{\Bc_b}/\Lm^\dl$.
With this identification, all the deformations are identified with vevs of
vector superfields: interacting for $D(\mu)>1$ and frozen for $D(\mu)\le0$.
\end{itemize}

\newpar{Relevance}
\nl There is another property of the deformation that can be deduced from the
dimension $D$ of the deformation parameter $\mu$: relevance. Marginal
deformations correspond to $D(\mu)=0$, relevant deformations --
to $D(\mu)>0$ and irrelevant -- to $D(\mu)<0$.
This distinction can also be identified in the worldsheet formulation%
\footnote{In the geometric formulation, for deformations $\dl W=\mu V$,
relevance means dominance of $V$ at $z\goto0$:
for a relevant deformation ($D(\mu)>0$), $r_V<1$ (see eq. (\ref{R-geo})),
so $V$ dominates over $W$ at $z\goto0$ and makes a macroscopic change in
the singularity;
for a marginal deformation ($D(\mu)=0$), $\mu$ parametrizes a continuous
change of the singularity; and for an irrelevant deformation, $V$ is
negligible at $z\goto0$ and has no effect on the singularity.}:
as described in the introduction, the coordinate $\phi$ parametrizes the
distance from the singularity. As in the holographic dualities with $AdS$
vacua, radial motion in the near-horizon geometry corresponds, in the
decoupled theory, to RG flow \cite{IMSY9802}
(see also \cite{SW9805}\cite{PP9808}):
\internote{Note that the Liouville field $\phi$ is identified with a scale
also in the context of 2D CFT coupled to quantum gravity
(see \cite{Seiberg90} for a review).}
large distances ($\phi\goto\infty$) correspond to high energies (UV) and
small distances ($\phi\goto-\infty$) -- to low energies (UV).
Therefore, a vertex operator (\ref{chiral}) with $\bt<0$
defines a perturbation that increases in the IR -- a relevant perturbation.
Similarly, $\bt<0$ corresponds to an irrelevant perturbation
and $\bt=0$, to a marginal one.
Since
\[ D(\mu)=-\frac{2\bt}{Q} \]
(see eqs. (\ref{DR}), (\ref{R-ws})),
these two approaches to identify relevance give the same result.

To summarize, we have made the identification $R''_+=R_+$
and it was used to reproduce, using the worldsheet formulation,
two bounds in the decoupled IR theory: the unitarity bound and relevance. 
The supercharges indeed satisfy $R''_+=R_+$, so the non-trivial content of
this identification is the charge of the vector superfields
(which are the only superfields discussed in this work).
Any one of the bounds could be used to {\em deduce} this equality.
Then, the agreement in all other aspects, as described above,
serves as additional evidence for the proposed holographic relation.

\newsection{IR Fixed Points in 4D $N=2$ SUSY Gauge Theories}
\secl{s-SGT}
In this section, we restrict attention to configurations relevant for the study
of 4D $N=2$ supersymmetric gauge theories. Specifically, we consider SQCD:
$SU(N_c)$ gauge theory with $N_f$ hypermultiplets (``quarks'')
in the fundamental representation of the gauge group.

\subsection{Realization of SQCD in String Theory}
\secl{s-brane}%
4D $N=2$ SQCD can be realized in string theory as follows
(for a review, see \cite{GK9802}):
one considers type IIA string theory with the following brane configuration:
all branes are extended in the (0123) direction; 
there are two NS5 branes which are also extended in (45),
and $N_f$ D6 branes which are also extended in (789);
finally, there are D4 branes extended also in the (6) direction over finite
intervals, ending on the other branes.
When there are $N_c$ coinciding D4 branes extending between the two NS5
branes, the low-energy dynamics of the D4 branes is described by 4D $N=2$
SQCD, as defined above \cite{HW9611}.
The vector multiplets correspond to open strings between the D4 branes,
and the hypermultiplets correspond to open strings between the D4 branes and
the D6 branes.
The dynamics of the NS5 and D6 branes is considered ``frozen'' in the 4D
theory, since these branes have infinite extension in the ``internal''
directions.
In particular, the $U(k)$ global (``flavor'') symmetry is the frozen gauge
symmetry of the D6 branes.
The relative location of these branes determines the parameters of the
gauge theory. In particular, the (45) locations of the D6 branes determine
the mass parameters of the hypermultiplets.
In a given NS5-D6 configuration, the possible locations of the D4 branes
(consistent with supersymmetry) parametrize the moduli space of the gauge
theory. In particular, distributing the D4 branes in the (45) directions
(along the NS5 branes), corresponds to the Coulomb branch, parametrized by
vacuum expectation values for the scalars in the vector multiplets.

Lifting this configuration to M-theory \cite{Witten9703}, the D6 branes
are identified as KK monopoles, corresponding to a non-trivial 4D transverse
space $\Mc$ (including the directions (456)) -- the multi Taub-NUT space
\cite{Hawk77}. One of the complex structures of $\Mc$ is that of a hypersurface
\beql{zwQ} zw=Q(x) \eeq
in $\CC^3$, where $Q(x)$ is a polynomial of degree $N_f$ with the coefficients
related to the mass parameters of the hypermultiplets.
In configurations corresponding to the Coulomb branch of the gauge theory,
the NS5 and D4 branes combine to a single M5 brane, wrapped on a 2D Riemann
surface $\Sg$, embedded holomorphicly in $\Mc$.
This surface is identified with the {\em Seiberg-Witten (SW) curve} of the
gauge theory \cite{SW9407}\cite{SW9408}%
\footnote{This curve was determined, using field-theoretic considerations,
in \cite{SW9407}\cite{KLTY9411}\cite{AF9411} (for pure $SU(N_c)$ SYM) and
\cite{SW9408}\cite{HO9505}\cite{APS9505} (for SQCD).}.
In the representation (\ref{zwQ}) of $\Mc$, $\Sg$ is the curve $H=0$, where%
\footnote{Substituting $w=Q(x)/z$ and $z=y+P(x)$ in $H=0$,
one obtains the familiar form
\[ y^2=P(x)^2-gQ(x) \hs. \]}
\beql{H-def} H=z+gw-2P(x) \hs. \eeq
Here $g$ is a function of the gauge coupling and $P(x)$ is a polynomial of
degree $N_c$, with the coefficients being the moduli of the Coulomb branch. 

Compactifying the (7) direction,
one obtains back type IIA string theory, this time with $N_f$ KK monopoles --
corresponding to the space $\RR^{3,1}\times\CC\times\Mc$ --
and an NS5 brane extended in $\RR^{3,1}$ and wrapped on $\Sg$.
This is a situation of the type considered in the introduction.
\internote{
As explained in section 2, although $\Sg$ is non-compact,
only a compact part of it contributes to its decoupled dynamics:
the region where $\Sg$ is highly curved (corresponding roughly to the
D4 branes in the first configuration described above).
This explains why the effective dynamics is 4-dimensional.}

Naively, the above relations seem to suggest that the decoupled dynamics on
this NS5 brane is described by $N=2$ SQCD.
This is not quite so, because the second description is valid in a range of
parameters (of string theory) which is different from that in which the gauge
theory was identified in the first configuration.
However, there is evidence that some aspects of the low energy dynamics
,including those studied below, are not sensitive to the above changes and,
therefore, are shared by the gauge theory and the dynamics of the NS5 brane.
Moreover, there is another chain of dualities relating gauge theories
(realized this time in heterotic string theory)
and the dynamics of NS5 branes in the same sense as above
(see \cite{Engineer} for reviews).

\subsection{Interacting IR Fixed Points in the Moduli Space}
\secl{s-IR}%
In most of the vacua in the Coulomb branch of $N=2$ SQCD, the massless
fields are vector superfields corresponding to an Abelian gauge symmetry and,
possibly, additional electrically-charged hypermultiplets. In these vacua,
the dynamics is free in the IR.
However, there are vacua with additional massless fields, for which the IR
dynamics is non-trivial, defining an interacting superconformal field theory
(SCFT).
This is the case for SQCD with $N_f=2N_c$ massless quarks, at the origin of
the Coulomb branch, where the additional massless fields can be identified,
at weak coupling, as vector superfields, enhancing the gauge symmetry to a
non-Abelian group.
Other non-trivial SCFT's are obtained at points in the Coulomb branch
in which mutually non-local hypermultiplets become massless \cite{AD9505}

All these vacua correspond to a SW curve $\Sg$ with an isolated singularity
so, in the stringy realization, there is an NS5 brane wrapped on a surface
$\Sg$ with an isolated singularity.
As explained in section \ref{s-dec}, the only part of the configuration that
is relevant for the decoupled dynamics on the NS5 brane is the neighborhood
of the singular point.
This implies that the relevant KK monopoles are those that are {\em at} the
singularity. With the singular point in $\Sg$ being at $x=0$, 
$Q(x)$ in (\ref{zwQ}) can be represented by  $Q(x)=x^k$ ($k\le N_f$),
corresponding to $k$ coinciding monopoles;
in the gauge theory this corresponds to $k$ massless quarks.
\internote{Remarks:
\nlb Singularity at $x\neq0$ would be interpreted as a non-vanishing vev,
canceled by a mass parameter.
\nls In this case, $x$ can be shifted to $0$, and then the above mass
parameter will appear as the coefficient of $x^{N_c-1}$ in $P(x)$.
\nlb The coefficient of $x^k$ in $Q(x)$ can be absorbed in $w$ and $g$,
reflecting the scale matching in the integration out of the massive quarks.}%
For the same reason, only the region near the center of the KK monopoles is
relevant and there the geometry is that of an orbifold $\CC^2/\ZZ_k$.
This orbifold can be parametrized by two flat coordinates $z',w'$,
subject to the identification
\beql{orb2} z'\goto\al z' \hsc w'\goto w'/\al \hsc \al=e^{2\pi i/k} \hs,
\eeq
and the coordinates $z,w,x$ in (\ref{zwQ}) are $\ZZ_k$-invariant functions
of $z',w'$:
\beql{zwx} z=z'^k \hsc w=w'^k \hsc x=z'w' \hs. \eeq

To summarize, we can consider type IIA string theory on
$\RR^{3,1}\times \CC^2_{z'w'}/\ZZ_k \times \CC$ with an NS5 brane on
$\RR^4 \times \Sg$, where $\Sg$ is the 2D surface $H(z,w,x)=0$ in
$\CC^2_{z'w'}/\ZZ_k$.
This has, as a dual description (see footnote \ref{f-T}),
type IIB string theory on $\RR^{3,1}\times X^6$,
where $X^6$ is a hypersurface $W=0$ in $\CC^2_{z'w'}/\ZZ_k\times\CC^2_{uv}$,
with the $\ZZ_k$ action (\ref{orb2}) and the polynomial
\beql{WH} W=H(z,w,x)+uv \hs. \eeq 

The surface $\Sg$ ($H=0$) has an isolated singularity at the origin and,
therefore, so does $X^6$ ($W=0$).
For a quasi-homogeneous $H$, this looks like a configuration of the type
considered in the previous section (with $d=4$, $n=3$ and $\Gm=\ZZ_k$).
Actually, there is a slight difference in the configurations but, as we shall
argue below, it is irrelevant for the study of the 4D gauge theory.
The difference is the following:
the orbifold considered in the previous section has a non-singular
worldsheet description and leads here to a $U(1)^k$ 6D gauge symmetry.
In the present construction, one has in mind an orbifold with an enhanced
$U(k)$ gauge symmetry, the extra massless states being related
(in the type IIA picture) to D2 branes wrapped on the vanishing cycles.
The difference in the backgrounds is that in the first one there is a
non-vanishing $B$-field, leading to non-vanishing masses for the above
wrapped D2 branes \cite{Asp9507}.
\internote{Remarks:
\nl In the orbifold CFT, the twisted sector consists of: [Witten9507, p.10]
\nls (c,c) ring $\goto$ complex deformations $\goto$ vectors (quark masses).
\nls (a,c) ring $\goto$ Kahler deformations (including the $B$-field)
$\goto$ hypers.
\nls vectors and hypers ``decouple'' in 4D $N=2$ SUSY.}
However, in the present context, \ie, concentrating on the 4D dynamics,
the 6D gauge symmetry on the orbifold is considered frozen anyway,
therefore, it is reasonable to expect that the above difference in
the 6D dynamics has no effect on the 4D theory.
Therefore, one can identify the present configuration as one of those
considered in the previous section.

\newpar{Study of the Interacting SCFT's}
\nl Using the above identification, one can use the anti-holographic
description proposed in section \ref{s-orb} to obtain information about
the interacting 4D SCFT's.
The general idea is to study deformations of the theory and through them,
to identify operators in it and to obtain information about their dynamics.
We concentrate, as in section \ref{s-orb}, on SUSY-preserving deformations,
defined in the worldsheet formulation by (c,c) operators of the
$LG_W/\ZZ_k$ CFT. 
As described in subsection \ref{s-four}, these are related to vector
superfields in the 4D theory and this relation depends on the dimension
$D=\half R_+$ of the corresponding deformation parameter $\mu$, as summarized
below:
\begin{itemize}
\item $D(\mu)>1$:

$\mu\sim\vev{\Ac_b}$, where $\Ac_b$ is the bottom component of a superfield
$\Ac$, whose dimension is $D(\Ac_b)=D(\mu)$;
the dimension $D>1$ indicates that this superfield is involved in a
non-trivial interaction;

\item $D(\mu)=1$ 

$\mu\sim\vev{\Bc_b}$, where $\Bc_b$ is the bottom component of a superfield
$\Bc$ which, in the IR, is frozen but couples to a non-trivial conserved
current;

\item $D(\mu)<1$:

the deformation is a change $\sim\mu\Ac$ in the prepotential of the 4D
SCFT, where $\Ac$ is an interacting vector superfield of dimension
$D(\Ac)=2-D(\mu)>1$;
$\mu$ is, possibly, related to a vev $\mu\sim\vev{\Bc}/\Lm^\dl$ of a free
superfield $\Bc$, which does not couple to a conserved current in the
IR SCFT and its only interaction with this theory is through an irrelevant
term $\Ac\Bc/\Lm^\dl$ in the superpotential (where $\Lm$ is a scale in the
underlying 4D theory and $\dl=1-D(\mu)>0$).
\end{itemize}
The deformations with $D(\mu)\neq1$ are expected to appear in pairs
($\mu_m,\mu_c$), each pair corresponding to an interacting vector
superfield $\Ac$, with the conformal dimensions related by
\beql{RAmu} D(\Ac)=D(\mu_m)=2-D(\mu_c) \hs. \eeq 

Some of this analysis can be performed directly in the gauge theory.
Indeed, for a deformation of the polynomial $W$, the dimension of the
deformation parameter can be obtained from eq. (\ref{R-geo}): 
\beql{Dmu} D(\mu)=\frac{1}{2}R'_+(\mu)=\frac{r_\mu}{r_\Om}=
\frac{1-r_V}{r_\Om} \hs, \eeq
thus using only geometric information about $X^6$:
the polynomial $W$ and the R-charge of the holomorphic 3-form $\Om$.
This information can be translated to purely gauge-theoretical data:
the deformations of $W$ (\ref{WH}) are the deformations of the SW
curve and the SW differential can be expressed as an integral over the
holomorphic 3-form $\Om$ \cite{KLMVW9604}, so they have the same R-charge.
Therefore, one can use the SW theory to find dimensions of deformation
parameters. This was indeed done in
\cite{AD9505}\cite{APSW9511}\cite{EHIY9603}.
However, the parametrization of the SW curve is not unique: for example,
the polynomials $P(x)$ and $Q(x)$ could be parametrized either by their
coefficients or by their roots, leading to a different set of conformal
dimensions. The embedding in string theory and the explicit worldsheet
formulation fixes this ambiguity (by identifying the deformation parameters
with coefficients of terms in the worldsheet Lagrangian)%
\footnote{As will be seen below, for the polynomials $P(x),Q(x)$,
the correct parameters are the coefficients of $P(x)$ and the roots of
$Q(x)$.}.
This demonstrates the usefulness of the holographic relation for the study of
these IR SCFT's.

We now analyze several families of SCFT's, using the worldsheet formulation
and compare the results to information from field theory.
Note that, in all cases, the superfields $u,v$ in eq. (\ref{WH}) decouple
(they are massive), so $LG_W=LG_H$.

\subsection{Singular Points in SYM}
\secl{s-SYM}%
For a singularity away from KK monopoles,
one can set $Q=1$ in eq. (\ref{zwQ}).
These are the singularities appearing in pure $N=2$ $SU(N_c)$ SYM theories%
\footnote{Unlike the singularities discussed in the next subsection, these
singularities appear only in the strong coupling region of the Coulomb moduli
space and do not have a semi-classical interpretation.}.

In this case, $g$ in eq. (\ref{H-def}) can be rescaled to 1 (by rescaling
the coordinates $z,w,x$ and the parameters in $P(x)$).
This is a reflection of the fact that the gauge coupling in the pure
SYM theory is transmuted to a scale and is not a real parameter in the
theory.
\internote{Details: $H=\sqrt{g}[z/\sqrt{g}+\sqrt{g}w-2P(x)/\sqrt{g}]$.}
The singularity appears for $P-1\sim x^l$ (with $2\le l\le N_c$).
The singular point is at $z=w=1$, and expanding around it
(with $z=1+\tilde{z}$), one obtains
\beql{H-pure} H\sim \tilde{z}^2-2(P-1)\sim\tilde{z}^2-2x^l \hs. \eeq
In the worldsheet formulation, the $LG_H$ CFT is a minimal model.
The (c,c) ring is spanned by 
\[ V=x^{l-j} \hsc  j=2,\ldots,l \hs. \]
Applying eq. (\ref{r-Om}) to (\ref{WH}),(\ref{H-pure}),
one obtains $r_\Om=1/2+1/l$,
\internote{The holomorphic form is $\Om=d\tilde{z}\wedge dx\wedge du/u$.}
which gives \cite{EHIY9603}\cite{GKP9907}, for the spectrum of conformal
dimensions (using eq. (\ref{Dmu})):
\[ r_\mu =jr_x=\frac{j}{l} \hs\hs\goto\hs\hs D(\mu)=\frac{2j}{l+2} \hs. \]
\internote{In the gauge-theory approach:
\nlb The curve is $y^2=(P-1)(P+1)$ and at the singularity ($P-1=x^l$):
\nl $y^2\sim2(P-1)=2x^l$, which is manifestly the same as in the CY approach.}
\internote{Flows: $l'\le l-2$.}
This spectrum corresponds to $[\half(l-1)]$ interacting vector superfields
(where $[\ldots]$ denotes the integer part):
\beql{D-pure}
\Uc_j \hsc D=\frac{2j}{l+2} \hsc j=\left[\frac{l}{2}+2\right],\ldots,l
\eeq
and, for even $l$, an additional IR-free vector superfield (with $D=1$).
\internote{The resulting spectrum of dimensions of couplings agrees with
the prediction of \cite{AD9505}, if they meant {\em anomalous} dimensions.}

Choosing $l=N_c$ (\ie, the most singular point for a given $N_c$),
one finds a nice agreement with the SW theory:
this singularity occurs at a singe point in the moduli space, which is
$(l-1)$-dimensional, so one expects to find $l-1$ different relevant
deformations of the IR SCFT.
Indeed, the present approach leads to $l-1$ deformations and all of them are
relevant: either vevs or relevant couplings ($D(\mu)<0$).
\internote{A puzzle: This seems not to work for $N_c>l$:
\nl The $x^{l-1}$ deformation seems to exist
in SW theory, but not in the stringy embedding.}
The SW theory also provides the charges (under the unbroken $U(1)^{l-1}$
gauge group) of the dyons that become massless at the singular point
\cite{SW9407}. In the present case one finds that (for an appropriate choice
of duality frame) out of the $l-1$ gauge fields, $[\half(l-1)]$ couple to
mutually-non-local dyons and, for even $l$, there is one more gauge field
with an electrically-charged hypermultiplet.
\internote{Details:
\nlb $2l$ branch points; $l$ of them ($i=1,\ldots,l$) coincide.
\nlb Choice of cycles: $\al_i=(2i-1,2i)$ (electric), $\bt_i=(2i,2i+1)$
(dyonic), $i=1,\ldots,l-1$.}
This is in full agreement with the present results:
the interacting superfields $\Uc_j$ are those coupled to mutually-non-local
dyons and the free superfield with $D=1$ is the one coupled only electrically.

The case $l=2$ is special, since it corresponds to a free IR SCFT.
Indeed, this singularity indicates in the gauge theory the appearance of a 
massless dyon. The low energy theory is SQED: a $U(1)$ vector multiplet
with an electrically charged hypermultiplet%
\footnote{This vector multiplet is related to the UV vector multiplet by a
duality transformation \cite{SW9407}.}.
There is a single deformation of this theory: a change of the vev of the
above vector superfield, giving a mass to the dyon.
Because of the electric charge, the $U(1)$ gauge coupling flows to zero in
the IR, so the vector superfield is frozen in the IR and, correspondingly,
the deformation parameter has $D=1$.
This is exactly what was found in the worldsheet formulation.

\subsection{Singular Points in SQCD}
\secl{s-SQCD}%
We turn to singularities with a non-trivial flavor symmetry ($k\ge2$).
The singular point is on the orbifold singularity $z'=w'=0$ and, neglecting
sub-leading terms, one obtains
\beql{Hlk} H=z+gw+x^l=z'^k+gw'^k-2h(z'w')^l \eeq
(corresponding to $P(x)\sim hx^l$).
We now distinguish between three situations:
\begin{itemize}
\item $k<2l$

In this case the last term is sub-leading, so $H$ in eq. (\ref{Hlk}) becomes
\beql{Hk} H=z'^k+gw'^k \hs. \eeq
Note that this is independent of $l$, therefore, so is the physics
at this singularity.
The $LG_H$ CFT is a product of two decoupled minimal models (they become
coupled by the $\ZZ_k$ orbifold action).
In the untwisted sector, the (c,c) ring is spanned by $\ZZ_k$-invariant
polynomials in $z',w'$ (which is the same as arbitrary polynomials in
$z,w,x$), with identifications modulo $\pt_{z'}H,\pt_{w'}H$.
A possible choice for a basis is
\[ V_j=x^j \hsc j=0,\ldots k-2 \hs. \]
Applying eq. (\ref{r-Om}) to (\ref{WH}),(\ref{Hk}),
one obtains $r_\Om=\frac{2}{k}=r_x$,
\internote{The holomorphic form is $\Om=dz'\wedge dw'\wedge du/u$.}
which gives \cite{EHIY9603}, for the spectrum of conformal dimensions
(using eq. (\ref{Dmu})):
\[ r_\mu =\left(\frac{k}{2}-j\right)r_x \hs\hs\goto\hs\hs
D(\mu)=\frac{k}{2}-j \hs. \]
The twisted sector of a LG orbifold was analyzed in \cite{IV90}.
Applying this analysis to the present case, one finds in each sector
one state $V'_i$ ($i=1,\ldots,2N-1$) with a $U(1)$ charge $f_{V'_i}=1-r_\Om$,
corresponding to $ D(\mu)=1$.
\internote{More details (from \cite{IV90}):
\nlb The charge of the RR vacuum (given in eq. (3.2)) vanishes,
since the contributions of $z'$ and $w'$ cancel each other.
\nlb The corresponding (c,c) state has, therefore $f=c/6=1-r_x$.
\nlb The orbifold action on this state is given in eq. (3.21).
\nls The discrete torsion is trivial and det($g$)=1;
it is possible to choose a trivial $K$ and with this choice,
the above state is invariant.
\nls Non-trivial $K$ is possible only for even $k$ and then the odd twisted
sectors are projected out.}
This leads to the identification of $[\half(k-2)]$ vector superfields
(where $[\ldots]$ denotes the integer part)
\beql{Dk}
\Uc_D \hsc \left\{\begin{array}{ll}
D=2,3,\ldots,\frac{k}{2}, & k \hbox{ even} \\
D=\frac{3}{2},\frac{5}{2},\ldots\frac{k}{2}, & k \hbox{ odd}
\end{array}\right.
\eeq
and $2[k/2]$ vector superfields with $D=1$.
\item $k=2l$

In this case, $H$ in eq. (\ref{Hlk}) is quasi homogeneous and, for $g\neq h$,
the surface $H=0$ has an isolated singularity
(the case $g=h$ will be discussed below).
As in the previous case, the $LG_H$ CFT is a product of two minimal models,
the only difference being that they interact through the last term in $H$.
This difference, however, has no effect on the chiral ring, which is,
therefore, the same as for $k<2l$.

\item $k>2l$

In this case, $H$ is not quasi-homogeneous. Naively, at least one of the
first two terms is negligible (\ie, has a higher weight), however,
neglecting it would lead to a non-isolated singularity (at $z'=0$ or $w'=0$).
A closer look reveals that the first term is leading
in the direction $w'=0$ and the second, in the direction $z'=0$, 
so the precise statement is that this kind of singularity cannot be described
by a quasi-homogeneous polynomial and, therefore, does not have a worldsheet
description of the form discussed in section \ref{s-orb}.
Note, however, that it can be realized indirectly, as a deformation of
$k\le2l$ singularities
\end{itemize}
\internote{Flows: Starting with $(l,k)$, $2l\ge k\ge2$,
\nlb $P\sim x^{l'}$ with $l'\le l-1,k-2$ leads to a $(l',k)$ singularity;
\nlb $Q$ deformation can lead to a $(l,k')$ singularity with $k'\le k-1$.}

\newpar{Comparing to Field Theory}
\nl Each of the above singularities appears in SQCD with $l\le N_c$,
$k\le N_f$, where $k$ of the quarks have the same mass parameter.
Moreover, each such singularity can be found in the {\em semi-classical
region} of an asymptotically-free theory (with $N_f<2N_c$ and $l<N_c$),
\ie, with mass parameters and vevs much larger then the strong coupling scale.
There, using semi-classical considerations, it can be identified
as the IR limit of $SU(l)$ gauge theory with $k$ massless quarks, at the
origin of the moduli space.

Using this identification, one can demonstrate that the holographic duality
considered here is a strong-weak coupling duality, in the sense that there
is no situation in which both the gauge theory and the worldsheet CFT
are weakly coupled.
For $k<2l$, the $LG_H$ theory is a product of two
decoupled minimal models, so the worldsheet CFT is solvable, but the gauge
theory is asymptotically free and strongly coupled in the IR.
For $k>2l$, the gauge theory is free in the IR but the $LG_H$ CFT is
complicated.
Finally, for $k=2l$ the situation depends on the interaction strength of the
last term in $H$ (in eq. (\ref{Hlk})). It is proportional to $h/\sqrt{g}$
(as can be seen by rescaling $w'\goto w'/g^{1/k}$),
so the minimal models are decoupled for $h/\sqrt{g}\goto0$ while the gauge
theory is free in the opposite limit (as will be described below).

As noted above, the singularity with $k<2l$ is independent of $l$. Moreover,
for even $k$ it is the same as $k=2l$ with $h\goto0$. This last relation can
be understood also in field theory, considering SQCD with
$N_f=k<2l<2N_c$ with even $k$, where the $SU(N_c$) gauge group is broken to
$SU(l)$ at some scale $M$ and further to $SU(l')$ ($l'=\half k$) at a lower
scale%
\footnote{To be able to apply semi-classical consideration, as is done above,
both scales should be sufficiently large, compared to the scale representing
the $SU(N_c)$ gauge coupling.}
$M'$. This gives the $(k,l')$ singularity, with
\[ h\sim \left(M^{N_c-l} M'^{l-l'}\right)^2 \hs. \]
Unbroken $SU(l)$ gauge group (corresponding to the $(k,l)$ singularity)
is obtained in the limit $M'\goto0$, which indeed implies $h\goto0$.

Therefore, there are two families of singularities, each labeled by
$l$, with $k=2l$ ($l\ge1$) and $k=2l-1$ ($l\ge2$) respectively%
\footnote{The singularity with $k=2l=2$ is the same as the $l=2$ singularity
in SYM and, as there, it corresponds to SQED (see the end of the previous
subsection).}.
Choosing $N_c=l$, the singularity appears at a single point in the
moduli space, which is $(l-1)$-dimensional, so, as in the SYM case, one
expects $l-1$ relevant deformations of the IR SCFT, and this is what is
found. For even $k$, these are the vevs of the $\half(k-2)=l-1$ vector
superfields in (\ref{Dk}), while for odd $k$, these are the $[\half(k-2)]=l-2$
vevs and one relevant coupling (that with $D(\mu)=\half$).
The distinction between couplings and vevs indicates that, for even $k$,
all the $l-1$ massless vector superfields are interacting, while for
odd $k$, one becomes free in the IR.

In addition, there are $k$ mass parameters for the quarks in the gauge theory.
As explained in subsection \ref{s-brane},
these can be identified with vevs of the vector superfields in the
Cartan subalgebra of the 6D $U(k)$ gauge theory on the coinciding KK monopoles
and these vector superfields are frozen in the 4D theory because of the
infinite extent of the KK monopoles in two additional (internal) directions.
These vector superfields couple to conserved (flavor) currents,
so they should correspond to deformations with $D(\mu)=1$.
Strictly speaking, only the $k-1$ non-diagonal flavor currents
(those in $SU(k)$) have the above interpretation.
The deformations corresponding to these currents are those found above in
the twisted sectors.

The diagonal flavor current has a somewhat more complicated nature.
If the gauge group was $U(N_c)$ instead of $SU(N_c)$, the diagonal flavor
current would couple to the $U(1)$ factor.
This is reflected also in the stringy embedding.
In fact, the original brane configuration, that with D4 branes,
realizes classically a $U(N_c)$ gauge theory, but it was argued in
\cite{Witten9703} that in this stringy realization, the vector superfield
corresponding to the diagonal $U(1)$ factor is frozen by quantum effects.
In the configuration with NS5 branes and KK monopoles,
the modulus of the $U(1)$ factor in 6D $U(k)$ gauge group corresponds to a
collective motion of all the KK monopoles in the $x$ direction,
while the corresponding modulus in the 4D $U(N_c)$ gauge group corresponds
to such a motion of the NS5 brane.
Since the quark masses depend on the relative displacement between the
KK monopoles and the NS5 brane, the diagonal flavor current couples to a
diagonal subgroup of the above two $U(1)$ factors.
For odd $k$, we did not find a deformation corresponding to this current.
This suggests that it decouples from the IR SCFT.
For even $k$, the corresponding deformation is the $D=1$ deformation from
the untwisted sector.
Indeed, the deformation is $\dl W\sim x^{l-1}\sim\pt_x W$,
which is an $x$-translation, corresponding to a relative motion
between the NS5 brane and the KK monopoles.

Additional deformations in the field theory are moduli of the Higgs branch,
corresponding to vevs of quark scalar fields. We do not find these in the
worldsheet description. Their absence can be understood as follows%
\footnote{This explanation was suggested to the author by D. Kutasov.}:
at the singular point, to which the Higgs branch is connected,
the quarks are massless.
This corresponds, in the geometric formulation, to the singular hypersurface
$X^6$, and in the worldsheet formulation, to a superpotential independent
of the Liouville superfield $\hat{\phi}$.
As explained in subsection \ref{s-pair}, in this situation the worldsheet
formulation suffers from a strong coupling singularity ($g_s\goto\infty$ for
$\phi\goto-\infty$) and to avoid this singularity, one should consider
the double scaling limit. In this limit, the mass $m$ of the quarks is kept
non-zero, which means that one actually considers a point in the Coulomb
branch {\em near} the singular one. By taking the limit $m\goto0$, one can
obtain information about the theory at the singularity, as was done above,
but clearly this limit will miss aspects that appear discontinuously only at
the singularity.
This is why one does not see the Higgs moduli in this approach.

\newpar{$SU(l)$ with $k=2l$ Massless Quarks}
\nl The $SU(l)$ gauge theory with $k=2l$ massless quarks is a
conformally-invariant theory and the (complexified)
gauge coupling $\tau$ is a modulus of the theory, which can be changed
continuously. In particular, for $\tau\goto i\infty$ one obtains
a free field theory (at all scales).
Therefore, unlike the other non trivial SCFT's, the one with $k=2l$ is
continuously related to a free theory, and the information implied by this
relation can be compared to that obtained from the stringy embedding.
This theory has an $SL(2,\ZZ)$ duality, relating different values of $\tau$.
The parameter $g$ in (\ref{H-def}) is an $SL(2,\ZZ)$-invariant function of 
$\tau$ (see \cite{HO9505}\cite{APS9505} for more details) and is, therefore,
the quantity characterizing the strength of the interaction.
Vanishing interaction corresponds to $g\goto0$, while for non
vanishing $g$, the theory is an interacting SCFT.

The origin of the moduli space corresponds
to $P(x)=x^l$ in (\ref{H-def}), so $h=1$ in (\ref{Hlk}).
For $g=1$, $W$ takes the form
\[ W=(z'^N-w'^N)^2+uv \hs, \]
which has a (non-isolated) $\ZZ_2$ singularity at $z'^N=w'^N$.
The corresponding NS5 brane configuration includes two coinciding NS5 branes,
which are the T-duals of the above $\ZZ_2$ singularity.
Thus, in this case, the decoupled dynamics includes a six-dimensional sector.
\internote{The NS5 configuration -- more details:
\nlb The simplest description is obtained in the coordinates $(z,x)$
that resolve the middle vanishing 2-cycle of the orbifold.
\nlb for generic $g$, there are $N$ NS5's wrapped on the vanishing cycle
$(x=0)$ and 2 more orthogonal to the others (constant $z$).
\nlb At $g=1$, the 2 NS5's coincide.}
\internote{Interpretation of the $g=1$ geometry:
\nlb Seems as a 4D intersection of $\ZZ_2$ and $\ZZ_N$ singularities.
\nl $\goto$ (naively) 4D $SU \times SU$.
\nls Instead, one finds 6D dynamics!
\nlb Also in [Katz,Mayr\&Vafa9706], the realization of $SU \times SU$
is dual to a IIA NS5 wrapped on a surface, which is different from
an intersection of $A_n$ singularities.
\nlb Since 2 coinciding flat NS5's lead to 6D dynamics, why the same is not
true for the KK monopoles?
\nls Could it be that they are frozen because we implicitly have a non-zero
$B$-field giving mass to the wrapped branes?}

To obtain an isolated singularity we, therefore, assume $g\neq1$.
In fact, we would like to consider $g\ll1$, where one can apply semi-classical
considerations. Then, the vector superfield $\Ac$ in the adjoint
representation defines the following gauge-invariant vector superfields
\[ \Uc_j=\tr\Ac^j \hsc j=2,\ldots,l \]
and their dimension is, classically (\ie, for $g\goto0$), $D(\Uc_j)=j$.
The gauge coupling $\tau$ is identified as the coefficient of a term
tr$\Ac^2$ in the prepotential.

This information from the gauge theory agrees with the results obtained
above from the stringy realization. The conformal dimensions found there
were independent of the coupling, so should be the same as for $g\goto0$
and, indeed, the correct spectrum is obtained (see eq. (\ref{Dk})).
The deformations $\dl H=\mu x^j$ correspond,
for $j\le l-1$, to a vev $\mu\sim\tr\vev{\Ac^{l-j}}$ and for $j\ge l$, to
a term $\mu\tr\Ac^{2+j-l}$ in the prepotential.
In particular, for $j=l$ (corresponding to a term $\mu\tr\Ac^2$), $\mu$
is a change in the gauge coupling%
\footnote{Note that such a marginal coupling (with $D(\mu)=0$) does not
appear in the other cases considered above, as expected from field theory.}.
This can also be seen directly in the polynomial $H$
(eq. (\ref{Hlk}) with $h=1$):
the corresponding deformation is $\dl H = \mu x^l$ and by rescaling $w'$,
this can be transformed to a change in $g$, which indeed represents the
gauge coupling.

\newsection{Summary}
\secl{s-sum}
In this work, type II string theory in a background of the form
\beql{LDbs} \RR^{d-1,1}\times\RR_\phi\times U(1)_Y \times\Cc \eeq
was interpreted as a $d$-dimensional theory. For $\Cc=LG_W$ (a 2D $N=2$
Landau-Ginzburg SCFT), this $d$-dimensional theory was identified in
\cite{GKP9907} as the description of the decoupled dynamics near an isolated
singularity in type II string theory on
\beql{CYbs} \RR^{d-1,1}\times X^{2n} \hs, \eeq
where $X^{2n}$ is a hypersurface $W=0$ in a flat space $\CC^{n+1}$
(with $d+2n=10$). Here this identification was extended to $\Cc=LG_W/\Gm$
(a Landau-Ginzburg orbifold), in which case, $X^{2n}$ is a hypersurface in
an orbifold $\CC^{n+1}/\Gm$. Furthermore, for $d=4$, $\Gm=\ZZ_k$ and special
choices of the polynomial $W$, this four-dimensional theory was related to
an interacting SCFT appearing in the moduli space of 4D $N=2$ SQCD: $SU(N_c)$
gauge theory with fundamental quarks.
Properties of the $d$ dimensional theory were identified in the worldsheet
formulation (of the string theory on (\ref{LDbs})) and contrasted, when
it was possible, with information from the geometric formulation
(of string theory on (\ref{CYbs}); in section \ref{s-orb})
and from the gauge theory (in section \ref{s-SGT}).
Here we summarize this analysis.

We start with properties that are independent of the details of $\Cc$
and depend only on $\Cc$ being a 2D CFT with (2,2) supersymmetry
and a spectral flow operator relating the (c,c) and (a,a) rings.
For any such $\Cc$:
\begin{itemize}
\item The theory has $2^{\frac{d}{2}+1}$ supercharges;

this corresponding to $N=2$ supersymmetry in 4 dimensions.
\item It has a $U(1)_+\times U(1)_-$ R-symmetry, under which all the
supercharges have charges $|R_\pm|=1$;

in $d=4$, where the R-symmetry group is $U(1)\times SU(2)$,
$R_+$ was identified with the $U(1)$ factor and $R_-$,
with a $U(1)$ subgroup of the $SU(2)$ factor;

furthermore, in the 4D IR SCFT, $R_+$ was identified as the $U(1)$ R-charge
appearing in the superconformal algebra (SCA)%
\ftl{f-id}{Analogous identifications can be made also for $d\neq4$.}.
\item Each operator in the (c,c) ring of $\Cc$ defines a deformation of
the theory, parametrized by a continuous parameter $\mu$ (which is a
coupling of a term in the worldsheet Lagrangian) with $R_-(\mu)=0$,
and this deformation preserves the supersymmetry of the theory;

in $d=4$, these deformations are related to scalars in vector multiplets
(which are, indeed, invariant under the $SU(2)$ R-symmetry):
$\mu$ is either a vev (modulus) of a bottom component $\Ac_b$ of a vector
superfield $\Ac$; or a coefficient (coupling) of a term in the 4D Lagrangian,
which is a top component $\Ac_t$ of such a superfield (corresponding to a term
$\mu\Ac$ in the prepotential);

this implies (using the SCA), that the conformal dimension of
the deformation parameter is$^{\ref{f-id}}$ (in both cases)
\beql{DRs} D(\mu)=\half R_+(\mu) \hs. \eeq
\item The value of $R_+(\mu)$ is related to the $\phi$-dependence of the
corresponding worldsheet vertex operator
(see eqs. (\ref{chiral}),(\ref{R-ws}))
and, consequently, provides information about the nature of the deformation:
for $R_+(\mu)>2$, this is a modulus and for $R_+(\mu)<2$, this is a coupling;
the coupling is relevant for $R_+(\mu)>0$, marginal for $R_+(\mu)=0$ and
irrelevant for $R_+(\mu)<0$;

in the IR CFT, the same information is provided by the conformal dimension
of $\mu$, where couplings are distinguished from moduli by using the
representation theory of the SCA (including the ``unitarity bound''
$D(\Ac_b)\ge1$ for a bottom components of a vector superfields $\Ac$);

using the identification (\ref{DRs}), these two approaches can be compared
and one finds identical distinctions.
\item Deformations with $R_+(\mu)\neq2$, were shown to appear in pairs;
for $d=4$ this is a coupling-modulus pairing and this is identified with
the pair of deformations that are defined by a given vector superfield.
\end{itemize}
Some of the evidence for the above identifications was obtained by considering
specific cases, as described below. However, it is natural to expect that
they have a more general range of validity.

We now turn to properties that are different for different $\Cc$ factors,
corresponding to differences between $d$-dimensional theories
(with $2^{\frac{d}{2}+1}$ supercharges).
First, considering $\Cc=LG_W/\Gm$ in general (including the case $\Gm=1$
studied in \cite{GKP9907}), the above identifications can be compared with
the geometric formulation of the theory, with the following results:
\begin{itemize}
\item One is led to the same requirements on the elements of $\Gm$.
\item One finds the same amount of supersymmetry:
$2^{\frac{d}{2}+1}$ supercharges.
\item The $U(1)_+$ R-symmetry is identified as a geometric isometry.
\item The (c,c) deformations from the untwisted sector are naturally
identified as complex structure deformations, induced by changes in the
polynomial $W$;
this identification was shown to be consistent with the multiplicity and
the R-charges of the deformation parameters.
\item For the above deformations, a distinction between parameters and
moduli can be identified also in the geometric formulation \cite{GVW9906}
and it was shown to agree with the distinction derived from the worldsheet
formulation.
\item Relevance of the above deformations translates, in the geometric
description, to dominance of the change in $W$ at $z\goto0$.
\end{itemize}

Finally, we considered specific four-dimensional SCFT's, appearing as
IR limits of SQCD, in singular points of its moduli space.
They are labeled by $(k,l)$, where $k$ is the number of massless quarks
at the singularity and $l$ is the degree of the singularity.
We considered three families of these singularities, corresponding to
$k=0,2l-1,2l$, and chose $l=N_c$, (corresponding to the most singular
point in the moduli space).
For these theories, there is independent information, from field-theoretic
considerations, and it was compared to the results form the worldsheet
analysis:
\begin{itemize}
\item The holographic duality was shown to be a strong-weak coupling
duality, in the sense that there is no situation in which both the gauge
theory and the worldsheet CFT are weakly coupled.
\item The effect of all the relevant and marginal (c,c) deformations
(in both the untwisted and twisted sectors) was identified in the
underlying gauge theory.
\item The deformations expected in the gauge theory are $N_c-1$
Coulomb moduli, $k$ mass parameters (couplings to the flavor currents) and,
for $k=2l$, a gauge coupling%
\footnote{The moduli of the Higgs branch are not expected, as explained in
subsection \ref{s-SQCD}.}.
The only deformation that was not identified is the coupling to the $U(1)$
flavor current for $k=2l-1$.
Its absence is an indication that this current decouples from the IR SCFT.
\item In pure SYM ($k=0$), field-theoretic considerations predict the
number of interacting superfields (with $D>1$), which are those coupled to
mutually-non-local charges, and the number superfields (with $D=1$) coupled
to (non-trivial) mutually local charges. This prediction agrees with the
worldsheet results.

As to the other families of SCFT's, the conformal dimensions (\ref{Dk})
indicate that, for $k=2l$, all the $N_c-1$ massless vector fields in the
Coulomb branch are interacting and, for $k=2l-1$, one of them is free.
\item For $k=2l$, the SCFT is continuously connected to a free theory by
changing the gauge coupling and the conformal dimensions
were found to be independent of such a change.
They should, therefore, be the same as in the free theory and indeed they
were found to be so.
\end{itemize}

The above detailed agreement between results obtained using different
approaches is a strong evidence for the duality proposed in \cite{GKP9907}
and in the present work. Additional evidence is found in the study of the
$d=6$ case, in \cite{GK9909}\cite{GK9911}.
This duality can now be used to study the decoupled theory on the singularity,
using the worldsheet formulation.
In particular, one can calculate correlation functions of observables in
the theory, as was done in \cite{GK9909}\cite{GK9911}.
This will provide information about 4D $N=2$ SCFT. It is also of interest to
extend this duality further, \eg, to other gauge groups and to theories
with less supersymmetry. All this is left for future study.

\vspace{1cm}
\noindent{\bf Acknowledgment:}
\nl I wish to thank my collaborators in this subject, A. Giveon and D. Kutasov,
for helpful discussions and exchange of ideas.
I am especially grateful to D. Kutasov, for numerous suggestion and comments.
I also benefitted from discussions with S. Kachru.
This work is supported in part by DOE grant \#DE-FG02-90ER40560


\appendix
\renewcommand{\newsection}[1]{
 \vspace{10mm} \pagebreak[3]
 \refstepcounter{section}
 \setcounter{equation}{0}
 \message{(Appendix \thesection. #1)}
 \addcontentsline{toc}{section}{
  App. \protect\numberline{\Alph{section}}{\hs\hs\boldmath #1}}
 \begin{flushleft}
  {\large\bf\boldmath Appendix \thesection. \hspace{5mm} #1}
 \end{flushleft}
 \nopagebreak}



\end{document}